\documentclass[aps, prd, twocolumn, nofootinbib, floatfix, superscriptaddress]{revtex4-1}
\usepackage{subfigure}
\usepackage{graphicx}
\usepackage{dcolumn}
\usepackage{epsfig}
\usepackage{amsmath}
\usepackage{amsfonts}
\usepackage{graphicx}
\usepackage{amssymb}
\usepackage{amssymb,amsmath,amsfonts}
\usepackage{xfrac}
\usepackage{color}
\usepackage{tikz}

\usepackage{tcolorbox}
\usepackage{hyperref}
\hypersetup{colorlinks, citecolor=bluscuro, linkcolor=black, urlcolor=bluscuro}
\definecolor{rossos}{cmyk}{0,1,1,0.55}
\definecolor{bluscuro}{rgb}{0.15, 0.2, .85}
\definecolor{bluchiaro}{cmyk}{1,.3,0.,0.1}

\newcommand{\arXiv}[2]{\href{http://arxiv.org/pdf/#1}{{\tt [#2/#1]}}}
\newcommand{\arXivold}[1]{\href{http://arxiv.org/pdf/#1}{{\tt [#1]}}}

\newcommand{\hr}{\hat{r}}

\newcommand{\p}{\prime}
\newcommand{\G}{\text{\tiny G}}

\newcommand{\be}{\begin{equation}}
\newcommand{\ee}{\end{equation}}
\newcommand{\bea}{\begin{eqnarray}}
\newcommand{\eea}{\end{eqnarray}}
\newcommand{\beq}{\begin{equation}}
\newcommand{\eeq}{\end{equation}}

\def\d{{\rm d}}
\def\hr{{\hat r}}

\def\G{\text{\tiny G}}

\newcommand{\no}{\nonumber}

\begin{document}
\title{ The Threshold for Primordial Black Hole Formation: a Simple Analytic Prescription  }

\author{ Ilia Musco }
\email{ilia.musco@unige.ch}
\affiliation{ \mbox{D\'epartement de Physique Th\'eorique and CAP, Universit\'e de Gen\`eve, 24 quai E. Ansermet, 
CH-1211 Geneva, Switzerland} }
\affiliation{ Instituto Galego de F\'isica de Altas Enerx\'ias, Universidade de Santiago de Compostela, 
E-15782 Santiago de Compostela, Spain } 

\author{ Valerio De Luca }
\email{valerio.deluca@unige.ch}
\affiliation{ \mbox{D\'epartement de Physique Th\'eorique and CAP, Universit\'e de Gen\`eve, 24 quai E. Ansermet, 
CH-1211 Geneva, Switzerland} }

\author{ Gabriele Franciolini }
\email{gabriele.franciolini@unige.ch}
\affiliation{ \mbox{D\'epartement de Physique Th\'eorique and  CAP, Universit\'e de Gen\`eve, 24 quai E. Ansermet, 
CH-1211 Geneva, Switzerland} }

\author{ Antonio Riotto } 
\email{antonio.riotto@unige.ch}
\affiliation{ \mbox{D\'epartement de Physique Th\'eorique and CAP, Universit\'e de Gen\`eve, 24 quai E. Ansermet, 
CH-1211 Geneva, Switzerland} }
\affiliation{INFN, Sezione di Roma, Piazzale Aldo Moro 2, 00185, Roma, Italy}

\begin{abstract} 
\noindent
Primordial black holes could have been formed in the early universe from non linear cosmological perturbations re-entering 
the cosmological horizon when the Universe was still radiation dominated. Starting from the shape of the power spectrum 
on superhorizon scales, we provide a simple prescription, based on the results of numerical simulations, to compute the 
threshold $\delta_c$ for primordial black hole formation. Our procedure takes into account both the non linearities between 
the Gaussian curvature perturbation and the density contrast and, for the first time in the literature, the non linear effects 
arising at horizon crossing, which increase the value of the threshold by about a factor two with respect to the one computed 
on superhorizon scales.
\end{abstract}

\maketitle


\section{Introduction and Summary}
\label{Introduction}
 It has been suggested that Primordial Black Holes (PBHs) might form in the radiation dominated era of the early Universe by 
 gravitational collapse of sufficiently large-amplitude cosmological perturbations~\cite{Zeldovich,Hawking,Carr1} (see 
 Refs.~\cite{sasaki, Green:2020jor} for recent reviews), and that they can comprise a significant fraction of the dark matter 
 in the universe, see Ref.~\cite{Carr:2020gox} for a review of the current experimental constraints on the PBH abundance. 
 This idea has recently received renewed attention given the possibility that PBHs might have given rise to gravitational 
 waves detected during the O1/O2 and O3 observational runs~\cite{LIGOScientific:2018mvr, LIGOScientific:2020stg, Abbott:2020khf, Abbott:2020tfl} 
 by the LIGO/Virgo Collaboration. This has motivated several studies on the primordial origin of these 
 events~\cite{Sasaki:2016jop,Bird:2016dcv,Clesse:2016vqa, Ali-Haimoud:2017rtz, raidal, raidalsm,ver, Gow:2019pok, 
 DeLuca:2020fpg, DeLuca:2020qqa, Clesse:2020ghq, Hall:2020daa, Jedamzik:2020ypm, Jedamzik:2020omx, DeLuca:2020sae, 
 DeLuca:2020jug}. In particular, the GWTC-2 catalog is found to be compatible with the primordial scenario \cite{1827880}.
 Furthermore, a possible detection of a stochastic gravitational wave background by the NANOGrav 
 collaboration~\cite{Arzoumanian:2020vkk} could be ascribed to PBHs~\cite{Vaskonen:2020lbd, DeLuca:2020agl, Kohri:2020qqd, 
 Domenech:2020ers, Sugiyama:2020roc,Inomata:2020xad}.
 
Despite some pioneering numerical studies~\cite{Nadezhin,Bicknell,Polnarev}, it has only recently become possible to fully
understand the mechanism of PBH formation with detailed spherically symmetric numerical simulations~\cite{Jedamzik:1999am,Shibata:1999zs,Hawke:2002rf,Musco:2004ak}, showing that a cosmological perturbation collapses 
to a PBH if it has an amplitude $\delta$ greater than  a certain threshold value $\delta_c$. This quantity has been estimated 
initially using a simplified Jeans length argument in Newtonian gravity~\cite{Carr2}, obtaining $\delta_c \sim c_s^2$, where 
$c_s^2=1/3$ is the sound speed of the cosmological radiation fluid measured in units of the speed of light. More recently, 
this value has been refined generalising the Jeans length argument with the theory of General Relativity, obtaining a value 
of $\delta_c\simeq0.4$ for a radiation dominated Universe~\cite{Harada:2013epa}. This analytical computation gives just a 
lower bound for the value of the threshold because it is not able to account for the non linear effects of pressure gradients, 
which require full numerical relativistic simulations. A recent detailed study has shown that there is a clear relation between 
the value of the threshold $\delta_c$ and the initial curvature (or energy density) profile, with $0.4\leq \delta_c \leq 2/3$, 
where the shape is identified by a single parameter~\cite{musco,Escriva:2019phb}.  

A consistent way to measure the amplitude of a perturbation is by using the relative mass excess inside the length scale 
of the perturbation, that for a consistent comparison between different shapes, should be measured at horizon crossing, 
when the length scale of the perturbation is equal to the cosmological horizon~\cite{musco}. 
  
Numerical simulations have also shown that the mechanism of critical collapse discovered by Choptuik~\cite{Choptuik:1992jv} 
is arising during the formation of PBHs, characterising the mass spectrum~\cite{Niemeyer:1997mt}. A crucial aspect to fully 
describe this mechanism was the implementation of an Adaptive Mesh Refinement (AMR), which allows study of the critical 
behavior down to very small values of $(\delta - \delta_c)$~\cite{Musco:2008hv,Musco:2012au}. 
 
Numerical simulations modelling PBH formation start from initial conditions specified on superhorizon scales, when the 
curvature perturbations describing adiabatic perturbations are time independent~\cite{Lyth:2004gb}. This allows expression
of the initial conditions of the numerical simulations, such as the energy density and velocity field, only in terms of a time 
independent curvature profile~\cite{Polnarev:2006aa,harada}, which can be derived, in the Gaussian approximation using 
peak theory~\cite{BBKS}, from the shape of the inflationary power spectrum of cosmological perturbations measured on 
superhorizon scales~\cite{Germani:2018jgr,Yoo:2018kvb}. 
 
 The relation between the shape of the peak of the curvature power spectrum and the initial conditions used in simulations 
 for PBH formation has recently been investigated in both the Gaussian approximation, with the aim of obtaining a proper 
 estimate of the cosmological abundance of PBHs~\cite{Young:2014ana,Germani:2018jgr,Yoo:2018kvb,Yoo:2020dkz}, and
 including also corrections coming from non linearities~\cite{ng2, ng4,DeLuca:2019qsy,Germani:2019zez,Young:2020xmk} 
 and non-Gaussianities~\cite{Young:2013oia,pbhng,Young:2015cyn, Yoo:2019pma, Kehagias:2019eil}. On the other hand, 
 numerical simulations have been used to reconstruct the shape of the peak of the inflationary power spectrum, understanding 
 to which extent this is consistent with the observational constraint for PBH formation on different scales~\cite{Kalaja:2019uju}.

 \subsection*{Prescription scheme} 
 \begin{tcolorbox}[colframe=white,arc=0pt,colback=gray!15
 ]
 
1. {\bf The power spectrum of the curvature perturbation}: take the primordial power spectrum $\mathcal{P}_\zeta$ of the  
Gaussian curvature perturbation and compute, on superhorizon scales, its convolution with the transfer function $T(k,\eta)$
\begin{equation*} 
P_\zeta(k,\eta) = \frac{2 \pi ^2}{k^3}\mathcal{P}_\zeta (k)  T^2(k,\eta).
\end{equation*}	

2. {\bf The comoving length scale $\hr_m$} of the perturbation is related to the characteristic scale $k_*$ of the 
 power spectrum $P_\zeta$. Compute the value of $k_* \hr_m$ by solving the following integral equation
 \[
\hspace{-0.45cm}\int  dk k^2 \!\left[ ( k^2\hr_m^2 - 1 ) \frac{\sin(k\hr_m)}{k\hr_m}  + \cos{(k\hr_m)} \right] \!P_\zeta(k,\eta) = 0\,. 
 \]
 
3. {\bf The shape parameter:} compute the corresponding shape parameter $\alpha$ of the collapsing perturbation, 
including the correction from the non linear effects, by solving the following equation
 \[
 \hspace{-0.5cm}F(\alpha) \left[ 1 + F(\alpha) \right] \alpha= \!- \!\frac{1}{2} \!\left[ 1 \!+\! \hr_m \!\frac{ \int dk k^4 \cos{(k\hr_m)} 
 P_\zeta (k,\eta) }{ \int dk k^3 \sin{(k\hr_m)} P_\zeta (k,\eta)} \right]
 \]
\[ F(\alpha) = \sqrt{ 1 - \frac{2}{5} e^{-\sfrac{1}{\alpha}} \frac{\alpha^{1-\sfrac{5}{2\alpha}}}{\Gamma\left(\frac{5}{2\alpha}\right) - 
\Gamma\left(\frac{5}{2\alpha},\frac{1}{\alpha}\right)} } \,. \]


4. {\bf The threshold $\delta_c$:} compute the threshold as function of $\alpha$, fitting the numerical simulations. 
\begin{itemize}
\item At \emph{superhorizon scales} making a linear extrapolation at horizon crossing ($aHr_m = 1$).
\[
\delta_c  \simeq
 \left\{
\begin{aligned} 
& \alpha^{0.047} - 0.50  \quad &0.1\lesssim \, \alpha \lesssim \ 7 \,  \\ 
& \alpha^{0.035}  - 0.475  \quad\quad  &7\lesssim \, \alpha \lesssim 13  \\ 
& \alpha^{0.026} - 0.45  \quad  &13\lesssim \, \alpha \lesssim 30
\end{aligned}
\right.
\]
\item At \emph{horizon crossing} taking into account also the non linear effects. 
 \[
\delta_c  \simeq
 \left\{
\begin{aligned}
 &\alpha^{0.125} - 0.05  \quad &0.1 \lesssim \, \alpha \lesssim 3 \\ 
&\alpha^{0.06} + 0.025  \quad\quad  & 3 \lesssim \,\alpha \lesssim 8 \\ 
& \quad \quad 1.15  \quad  &\alpha \gtrsim 8 
\end{aligned}
\right. 
 \]  
 \end{itemize}
 The difference between these two values of the threshold $\delta_c$ is discussed later in section \ref{non_linear_horizoncrossing}. 
\end{tcolorbox}

The aim of the present paper is to enable the interested reader to calculate the threshold for PBH formation, when the 
 Universe is still radiation dominated, without the need for running numerical simulations. Although non linear cosmological 
 density perturbations are described by a non Gaussian random field, we provide a simple prescription to compute the 
 threshold $\delta_c$ from the shape of the Gaussian inflationary power spectrum. The algorithm, divided into a few simple 
 steps, accounts for both the non linearities associated with the relation between the Gaussian curvature perturbation and 
 the density contrast as well as for those, so far neglected in the present literature, arising at horizon crossing. While a more 
 refined description of the various steps will  be found in the rest of the paper, we here provide the reader in the adjacent 
 column with an overview.

 Following the present Introduction, Section~\ref{initial conditions} reviews the mathematical formulation of the 
 problem. In Section~\ref{shape} we discuss the relation between the threshold $\delta_c$ and the shape of cosmological 
 perturbation. In Section~\ref{average} we show how to compute the typical value of the threshold $\delta_c$ as a function of 
 the shape of the power spectrum, analysing in detail some explicit examples. In Section~\ref{non_linear_horizoncrossing} 
 we compute, using numerical simulations, the amplitude of the threshold $\delta_c$ as a function of the shape parameter,
 discussing the difference when this is computed at horizon crossing or on superhorizon scales. Finally in Section~\ref{conclusions} 
 conclusions are presented, making a summary of the results. Throughout we use $c = G = 1$.


\section{Initial conditions for PBH formation}
\label{initial conditions}
\subsection{Gradient expansion}
\noindent

PBHs form from the collapse of non-linear cosmological perturbations after they re-enter the cosmological horizon. 
Following the standard result for extreme peaks we assume spherical symmetry on superhorizon scales~\cite{BBKS}.
The local region of the Universe characterised by such 
perturbations is described by an asymptotic form of the metric, usually written as
\bea
\d s^2 & = & -\d t^2 + a^2(t) \left[ \frac{\d r^2}{1-K(r)r^2} + r^2\d\Omega^2 \right]   \no \\
& = & - \d t^2+a^2(t)e^{2\zeta(\hr)}  \left[ \d\hr^2 + \hr^2\d\Omega^2 \right] ,
\label{pert_metric}
\eea
where $a(t)$ is the scale factor, while $K(r)$ and $\zeta(\hr)$ are the conserved comoving curvature perturbations 
defined on a super-Hubble scale, converging to zero at infinity where the Universe is taken to be unperturbed and 
spatially flat. The equivalence between the radial and the angular parts gives
\begin{equation} 
\left\{
\begin{aligned}
& r = \hr e^{\zeta(\hr)} \,,\\ 
& \displaystyle{\frac{\d r}{\sqrt{1-K(r)r^2}}} = e^{\zeta(\hr)} \d\hr \,,
\end{aligned}
\right.
\label{K_zeta}
\end{equation}
and the difference between the two Lagrangian coordinates $r$ and $\hr$ is related to the ``spatial gauge" of the 
comoving coordinate, which is fixed by the form chosen to specify the curvature perturbation put into the metric, i.e. 
$K(r)$ or $\zeta(\hr)$. From a geometrical point of view the coordinate $\hr$ considers the perturbed region as a 
local FRW separated universe, with the curvature perturbation $\zeta(\hr)$ modifying  the local expansion, while 
the curvature profile $K(r)$ is defined with respect to the background FRW solution ($K=0$). Combining the 
two expressions in~\eqref{K_zeta} one gets
\be 
K(r)r^2 =  - \hr\zeta'(\hr) \left[ 2+\hr\zeta'(\hr) \right],
\ee
showing that $K(r)$ is more directly related to the spatial geometry of the spacetime, obtained as a quadratic 
correction in terms of $\hr\zeta^\p(\hr)$. 
 
On the superhorizon scales, where the curvature profile is time independent, we use the gradient expansion 
approach~\cite{Shibata:1999zs,Tomita:1975kj,Salopek:1990jq,Polnarev:2006aa}, based on expanding the time 
dependent variables such as energy density and velocity profile, as power series of a small parameter $\epsilon \ll 1$ 
up to the first non zero order, where $\epsilon$ is conveniently identified with the ratio between the Hubble radius and 
the length scale of the perturbation. This approach reproduces the time evolution of linear perturbation theory but also 
allows having non linear curvature perturbations if the spacetime is sufficiently smooth on the scale of the perturbation
(see~\cite{Lyth:2004gb}). This is equivalent to saying that pressure gradients are small when $\epsilon \ll 1$ and are not 
then playing an important role in the evolution of the perturbation. 

In this approximation, the energy density profile can be written as  
\cite{harada,musco} 
\bea
\label{rel} 
\frac{\delta\rho}{\rho_b} &\equiv& \frac{\rho(r,t) - \rho_b(t)}{\rho_b(t)} =  
\frac{1}{a^2H^2} \frac{3(1+w)}{5+3w} \frac{ \left[ K(r)\,r^3 \right]^\prime}{3r^2} \no \\
&=& - \frac{1}{a^2H^2} \frac{4(1+w)}{5+3w} e^{-5\zeta(\hr)/2}\nabla^2 e^{\zeta(\hr)/2} \,,
\eea
where $H(t)=\dot{a}(t)/a(t)$ is the Hubble parameter, {\bf$\rho_b$ is the mean background energy density} and $K'(r)$ denotes differentiation with respect to $r$ while 
$\zeta'(\hr)$ and $\nabla^2\zeta(\hr)$ denote differentiation with respect to $\hr$. The parameter $w$ is the coefficient 
of the equation of state $p = w \rho$ relating the total (isotropic) pressure $p$ to the total energy density $\rho$. From 
now on we are going to consider just the standard scenario for PBH formation assuming a radiation dominated Universe 
with $w=1/3$.

\subsection{The compaction function}
\noindent
The criterion to distinguish whether a cosmological perturbation is able to form a PBH depends on the amplitude measured 
at the peak of the compaction function~\cite{Shibata:1999zs, musco} defined as
\be
\label{a}
\mathcal{C} \equiv 2\frac{\delta M(r,t)}{R(r,t)} \,,
\ee
where $R(r,t)$ is the areal radius and $\delta M(r,t)$ is the difference between the Misner-Sharp mass within a sphere 
of  radius $R(r,t)$, and the background mass \mbox{$M_b(r,t)=4\pi \rho_b(r,t)R^3(r,t)/3$} within the same areal radius 
but calculated with respect to a spatially flat FRW metric. In the superhorizon regime (i.e. $\epsilon \ll 1$) the compaction 
function is time independent, and is simply related to the curvature profile by
\be \label{comp}
\mathcal{C} = \frac{2}{3} K(r)r^2 =  - \frac{2}{3} \hr\zeta'(\hr) \left[ 2+\hr\zeta'(\hr) \right] \,.
\ee
As shown in \cite{musco}, the comoving length scale of the perturbation is the distance from $r=r_m$, where the 
compaction function reaches its maximum (i.e. \mbox{$\mathcal{C}'(r_m) = 0$}), which gives
\bea 
K(r_m)+\frac{r_m}{2}K'(r_m)=0, 
\eea
or
\bea
\label{rm_condition} 
\zeta'(\hr_m)+\hr_m\zeta''(\hr_m)=0 \,.
\eea
Given the curvature profile, the parameter $\epsilon$ of the gradient expansion is defined as
\be
\epsilon \equiv \frac{R_H(t)}{R_b(r_m,t)} = \frac{1}{aHr_m} = \frac{1}{aH\hr_m e^{\zeta(\hr_m)}} \,,
\ee
where $R_H=1/H$ is the cosmological horizon and \mbox{$R_b(r,t)=a(t)r$} is the background component of the areal 
radius. With these definitions, the expression written in Eq.~\eqref{rel} is valid for $\epsilon \ll 1$.

\subsection{The perturbation amplitude and the threshold}
We are now able to define consistently the perturbation amplitude as being the mass excess of the energy density within 
the scale $r_m$, measured at the cosmological horizon crossing time $t_H$, defined when $\epsilon=1$ ($aHr_m=1$). 
Although in this regime the gradient expansion approximation is not very accurate, and the horizon crossing defined in 
this way is only a linear extrapolation, this provides a well defined criterion to measure consistently the amplitude of different 
perturbations, understanding how the threshold is varying because of the different initial curvature profiles (see~\cite{musco} 
for more details). Later in Section~\ref{non_linear_horizoncrossing} we are going to extend the present discussion to include 
the non linear effect on the threshold when the cosmological horizon crossing is fully computed with numerical simulations.  

The amplitude of the perturbation measured at $t_H$, which we refer to as $\delta_m \equiv \delta(r_m,t_H)$, is given by 
the excess of mass averaged over a spherical volume of radius $R_m$, defined as
\be
\label{delta}
\delta_m = \frac{4\pi}{V_{R_m}} \int_0^{R_m}   \frac{\delta\rho}{\rho_b} \,R^2 \d R\,  = 
\frac{3}{r_m^3} \int_0^{r_m} \frac{\delta\rho}{\rho_b} \, r^2 \d r  \,,
\ee
where $V_{R_m} = {4\pi}R_m^3/3$. The second equality is obtained by neglecting the higher order terms in $\epsilon$, 
approximating \mbox{$R_m \simeq a(t)r_m$}, which allows to simply integrate over the comoving volume of radius $r_m$. 
Inserting the  expression for $\delta\rho/\rho_b$ given by \eqref{rel} into \eqref{delta}, one obtains 
$\delta_m = \mathcal{C}(r_m)$ and a simple calculation seen in~\cite{musco} gives the fundamental relation
\be
\delta_m = 3 \frac{\delta\rho}{\rho_b} (r_m,t_H) \,.
 \label{delta_m}
\ee

PBHs form when the perturbation amplitude $\delta_m > \delta_c$, where the value of the threshold $\delta_c$ depends 
on the shape of the energy density profile, with $2/5 \leq \delta_c \leq 2/3$, as shown in~\cite{musco}. Defining the quantity 
$\Phi \equiv -\hr\zeta^\prime(\hr)$ we can write $\delta_m$ as
\be \label{Phi}
\delta_m = \frac{4}{3} \Phi_m \left( 1 - \frac{1}{2} \Phi_m \right)  
\ee
where $\Phi_m = \Phi(\hr_m)$, and the corresponding threshold for $\Phi$ is such that $0.37 \lesssim \Phi_c \leq 1$. 

This shows that there are two different values of $\Phi_m$ corresponding to the same value of $\delta_m$, with a maximum 
value of $\delta_m=2/3$ for $\Phi_m=1$. This degeneracy in the amplitude of the perturbation measured with $\delta_m$ 
is related to the difference between cosmological perturbations of Type I and Type II that have been carefully analysed 
in~\cite{Kopp:2010sh}. Here we review this analysis in the context of PBHs formation. 

The quantity $\Phi_m$ measures the perturbation amplitude in terms of the local curvature, uniquely defined, 
while the quantity $\delta_m$ is related to the global geometry, related to the compactness of the region or radius $r_m$, 
which has a degeneracy: there are two possible geometrical configurations of the spacetime with the same compactness 
as shown in  Figure 3 of~\cite{Kopp:2010sh}. When $\Phi>1$ the spatial geometry of the spacetime starts to close on itself, 
up to $\Phi=2$ corresponding to the \emph{Separate Universe} limit. 

Computing the first and second derivatives of $\mathcal{C}$ in terms of $\Phi$ gives
\bea 
 \mathcal{C}^\prime(\hr) &=& \frac{4}{3} \Phi^\prime(\hr) \left( 1 - \Phi(\hr) \right), \\
  \mathcal{C}^{\prime\prime}(\hr) &=&  \frac{4}{3} \left[ \Phi^{\prime\prime}(\hr) \left( 1 - \Phi(\hr) \right) - \left( \Phi^\prime(\hr) \right)^2 \right].
  \label{C_2nd}
\eea
For a positive peak of the density contrast $\delta\rho/\rho_b$ we have $\mathcal{C}^\prime(\hr_m) = 0$ and 
$\mathcal{C}^{\prime\prime}(\hr_m) < 0$, and one can distinguish between PBHs of Type I and Type II from the sign 
\mbox{of $\Phi^{\p\p}_m$.}

\begin{itemize}

\item 
{\bf PBHs of Type I:} $\delta_c \!< \!\delta_m \leq 2/3$ and \mbox{$\Phi_c\! < \Phi_m\leq\! 1$.} \\
In this case $\delta_m$ is increasing for larger values of $\Phi_m$ and \mbox{$\mathcal{C}^\prime(\hr_m)=0$} implies 
that $\Phi^\prime_m=0$, corresponding to the condition for $\hr_m$ given in \eqref{rm_condition}. When $\Phi_m \leq 1$ 
we have $\mathcal{C}^{\prime\prime}(\hr_m) < 0$ corresponding to $\Phi^{\prime\prime}_m<0$. In the limiting case of 
$\Phi_m=1$ we have that both $\Phi^{\prime\prime}_m$ and $\mathcal{C}^{\prime\prime}(\hr_m)$ are converging towards  
$-\infty$ (see after \eqref{alpha_phi} for more explanations).

\item
{\bf PBHs of Type II:} $2/3 > \delta_m \geq 0$ and $1<\Phi_m\leq2$. \\
In this case $\delta_m$ is decreasing for larger values of $\Phi_m$, and as before \mbox{$\mathcal{C}^\prime(\hr_m)=0$} 
implies that $\Phi^\prime_m=0$ (see Eq.\eqref{rm_condition}). For $\Phi_m>1$ we have 
$\mathcal{C}^{\prime\prime}(\hr_m) < 0$ while $\Phi^{\prime\prime}_m>0$, changing sign with respect to \mbox{Type I} solutions. 

\end{itemize}  

All of the possible values of the threshold are within the regime of PBHs of Type I, where the mass spectrum of PBHs 
has a behavior described by the scaling law of critical collapse~\cite{musco} 
\be
\label{scaling}
M_\textrm{PBH} = {\cal{K}} (\delta_m - \delta_c)^\gamma M_H \,,
\ee
with $\gamma\simeq0.36$ for a radiation dominated fluid, where $M_H$ indicates the mass of the cosmological horizon 
measured at time $t_H$ and $\cal{K}$ is a coefficient depending on the particular profile of $\delta\rho/\rho_b$. Numerical 
simulations have shown that $1 \lesssim {\cal{K}} \lesssim 10$, and that~\eqref{scaling} is valid with $\gamma$ constant 
when $\delta_m-\delta_c \lesssim 10^{-2}$.


\section{The shape parameter}
\label{shape}
\begin{figure}[t!]
 \vspace{- 1.5cm}
 \centering
   \includegraphics[width=0.48\textwidth]{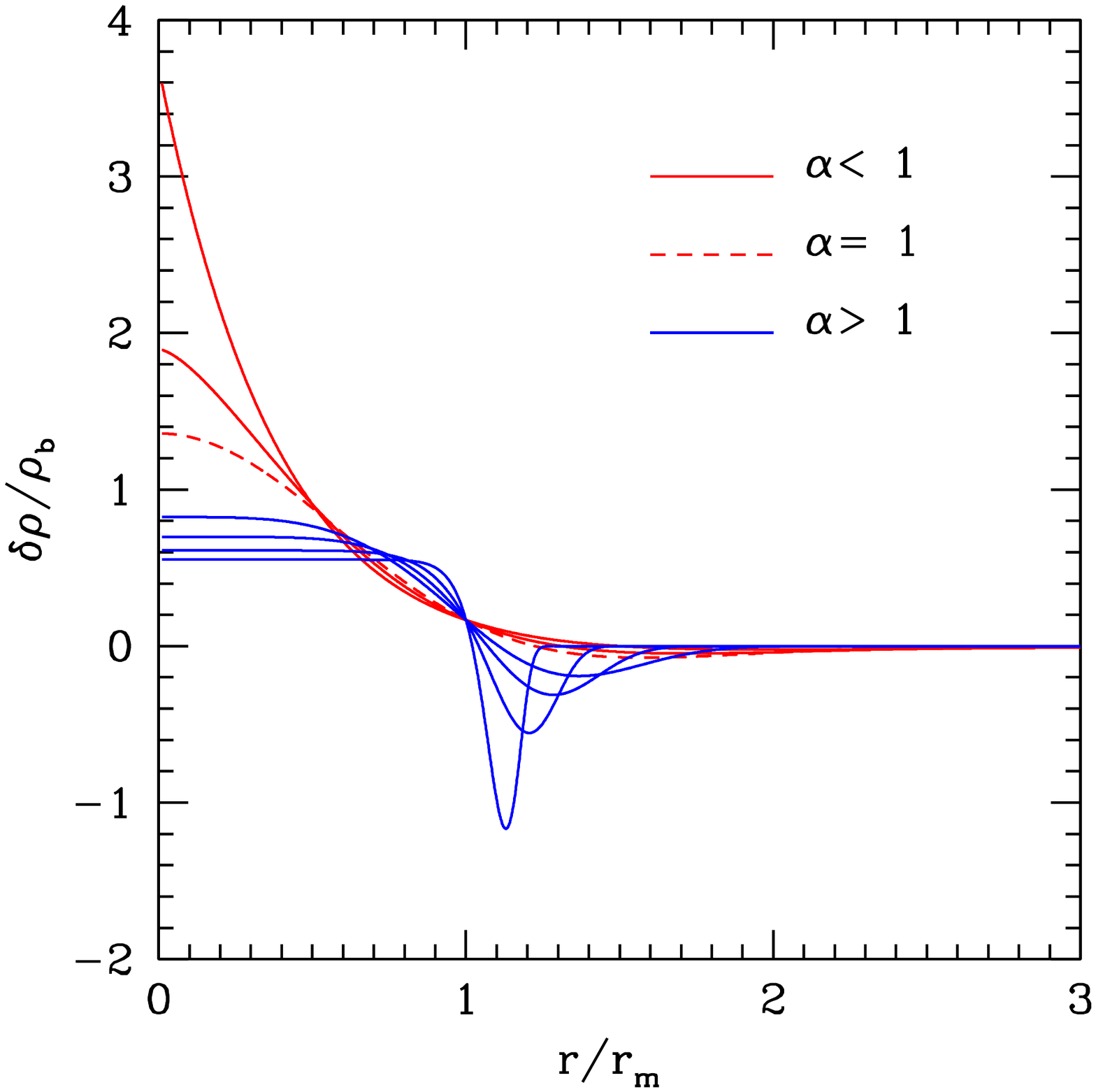} 
  \vspace{-2.5cm}
  \caption{ This figure, taken from~\cite{musco}, shows the behavior of $\delta\rho/\rho_b$ given by~\eqref{rel} plotted 
  against $r/r_m$ when \mbox{$aHr_m=1$}, for $\alpha=0.5, 0.75, 1, 2, 3, 5, 10$. The profiles with $\alpha\leq1$ are 
  plotted with a red line (a dashed line for $\alpha=1$) while blue lines are used for profiles with $\alpha>1$. } 
  \label{rho_shapes}
 \end{figure}

As seen in~\cite{musco,Escriva:2019phb}, the threshold for PBHs depends on the shape of the cosmological perturbation, 
characterised by the width of the peak of the compaction function, measured by a dimensionless parameter defined as
\be \label{alpha}
\alpha = - \frac{ \mathcal{C}^{\prime\prime}(r_m) r_m^2 }{ 4 \mathcal{C}(r_m) },
\ee
where the family of curvature profiles $K(r)$ given by
\be \label{K_basis}
K(r) = \mathcal{A} \exp \left[ -\frac{1}{\alpha} \left( \frac{r}{r_m} \right)^{2\alpha} \right]
\ee
identifies a basis of profiles which describes the main features of all of the possible shapes. In Figure~\ref{rho_shapes} - 
taken from~\cite{musco} - one can see the energy density profile $\delta\rho/\rho_b$ plotted against $r/r_m$, obtained 
by inserting~\eqref{K_basis} into~\eqref{rel} for different values of $\alpha$, normalised at horizon crossing ($aHr_m=1$). 
The shape of the energy density contrast becomes peaked for $\alpha<1$ (red lines) corresponding to a broad profile of 
the compaction function, where the dashed line describes the typical Mexican-hat profile ($\alpha=1$). On the contrary 
the shape of the compaction function $\mathcal{C}$ is more peaked for values of $\alpha>1$ (blue lines), corresponding 
to broad profiles of the density contrast.  

It is important to appreciate that when replacing~\eqref{K_basis} or any other $K$-profile into~\eqref{alpha}, the value of 
$\alpha$ is independent of the amplitude $\delta_m = \mathcal{C}(r_m)$ of the perturbation, related to the peak 
$\mathcal{A}$ of the curvature profile: this value is just an overall factor which cancels out in the ratio between the 
second derivatives and the value of the peak of the compaction function. The parameter $\alpha$ is therefore 
distinguishing between different shapes of the perturbation, independently of their amplitude.  

As shown in~\cite{Escriva:2019phb}, the average value of $\mathcal{C}(r)$ integrated over a volume of comoving 
radius $r_m$, defined as
\be
\label{delta_av}
\bar{\mathcal{C}}(r_m) =  \frac{3}{r_m^3} \int_0^{r_m} \mathcal{C}(r) \, r^2 \d r  \,,
\ee
has a nearly constant value when computed at the threshold for PBH formation, which is $\bar{\mathcal{C}}_c\simeq2/5$. 
This allows derivation of an analytic expression to compute the threshold $\delta_c$ as a function of the shape parameter 
$\alpha$, up to a few percent precision~\cite{Escriva:2019phb}
\be \label{delta_c}
\delta_c \simeq \frac{4}{15} e^{-\sfrac{1}{\alpha}} \frac{\alpha^{1-\sfrac{5}{2\alpha}}}{\Gamma\left(\frac{5}{2\alpha}\right) - 
\Gamma\left(\frac{5}{2\alpha}, \frac{1}{\alpha} \right)},
 \ee
 where $\Gamma$ identifies the special Gamma-functions. This is consistent with the analysis made in~\cite{musco} 
 where it was shown that the effects of additional parameters modifying the simple basis given by~\eqref{K_basis} 
 are negligible.
 
The corresponding peak amplitude $\delta\rho_0/\rho_b$, corresponding to the overdensity amplitude evaluated at 
the centre of symmetry, is related to the value of $\delta_m$ by $\delta\rho_0/\rho_b = e^{\sfrac{1}{\alpha}} \delta_m$, 
which combined with~\eqref{delta_c} gives
 \be \label{peak}
 \left(\frac{\delta\rho_0}{\rho_b}\right)_c \simeq \frac{4}{15} 
 \frac{\alpha^{1-\sfrac{5}{2\alpha}}}{\Gamma\left(\frac{5}{2\alpha}\right) - \Gamma\left(\frac{5}{2\alpha}, \frac{1}{\alpha} \right)}\,.
\ee
The shape parameter $\alpha$ describes the main features of the profile in the region $0 < r \lesssim r_m$ where  PBHs 
form, while any other additional parameters describe only secondary modification of the tail, $r \gtrsim r_m$, giving only a 
few percent deviation of the value of $\delta_c$ with respect to the one obtained with~\eqref{K_basis}. 

The shape is not correlated with the amplitude of the perturbation when the shape is measured in the $r$-gauge of the 
comoving coordinate, while a correlation arises when measured in the $\hr$-gauge. Using the coordinate transformation 
of~\eqref{K_zeta} one obtains that
\be 
\mathcal{C}^{\prime\prime}(r_m) = \frac{1}{ e^{2\zeta(\hr_m)} \left[ 1 + \hr_m\zeta(\hr_m) \right]^2 } 
\mathcal{C}^{\prime\prime}(\hr_m),
\ee
where the additional term proportional to $\mathcal{C}^{\prime}(\hr)$ is equal to zero when calculated
at $\hr_m$ because of \eqref{rm_condition}. As stated in the introduction, the prime denotes spatial derivative 
with respect to the variable written explicitly in the argument of the function.
The shape parameter can therefore be written as 
\be \label{alpha_zeta}
\alpha = -  \frac{ \mathcal{C}^{\prime\prime}(\hr_m) \hr_m^2 } { 4\mathcal{C}(\hr_m) 
\left[ 1 - \frac{3}{2} \mathcal{C}(\hr_m) \right] },
\ee 
showing that the peak of the compaction function does not cancel out with the peak of the second derivative, 
when computed with respect to $\hr$ instead of $r$. 

Using~\eqref{Phi}, this can be written as
\be \label{alpha_phi}
\alpha = - \frac{ \Phi_m^{\prime\prime} \,\hr_m^2 }{ 4\Phi_m \left(1 - \frac{1}{2}\Phi_m \right) \left(1-\Phi_m\right) },
\ee
showing that in general, when varying the amplitude of the perturbation, the values of $\Phi_m''$ and $\Phi_m$ are not 
independent, but correlated, changing according to the given value of $\alpha$. It is interesting to note that both 
Type I (\mbox{$ \Phi_m^{\prime\prime}\leq0, \Phi_m\leq1$}) and Type II ($ \Phi_m^{\prime\prime}>0, \Phi_m>1$) 
perturbations have $\alpha>0$, consistently with~\eqref{alpha}. 

In general there is a correlation between the shape of $\Phi$ and the amplitude of the peak $\Phi_m$. In the upper limit 
of the Type I solution, when $\Phi_m \to 1$, one finds $\alpha \to \infty$ which implies from~\eqref{C_2nd} that 
$\Phi_m^{\p\p} \to -\infty$ because $\mathcal{C}^{\prime \prime}(r_m) <0$ for any positive peak of the compaction function. 
From the geometrical point of view the shape of the compaction function is forced to be a Dirac delta (a top hat in the energy 
contrast) when $\Phi_m = 1$, corresponding to the threshold for PBH formation when $\alpha\to\infty$. 

To give an explicit example of the correlation between the amplitude and the shape of $\zeta(\hr)$ we can consider the 
profile used in~\cite{ng4}
\be
\zeta(\hr) = \mathcal{B} \exp \left[ - \left( \frac{\hr}{\hr_m} \right)^{2\beta} \right],
\ee 
that inserted into the~\eqref{alpha_phi} gives
\be
\alpha = \frac{ \beta^2 }{ (1-\beta\zeta(\hr_m))(1-2\beta\zeta(\hr_m)) } \,.
\ee
In the linear approximation $\mathcal{B}\ll1 \, \Rightarrow \, \beta\zeta(\hr_m)\ll1$, which gives $\alpha \simeq \beta^2$,  
showing that for a given value of $\alpha$, the corresponding value of $\beta$ is fixed and there is no correlation between 
the shape and the amplitude, while when we are considering a perturbation amplitude of the order of the threshold $\delta_c$, 
one has $\mathcal{B}\sim1$ (corresponding to \mbox{$\mathcal{A} r_m^2\sim1$}) and the correlation is not negligible. 
For example, when $\alpha=1$ one has a typical Mexican-hat shape and a value of the threshold $\delta_c \simeq 0.5$, 
while for a value of \mbox{$\beta=1$} corresponding to a Mexican-hat shape in the linear approximation, the value of the 
threshold is \mbox{$\delta_c\simeq0.55$}, as seen in~\cite{ng4}.  

\begin{figure*}[t!]
\vspace{-1cm}
 \centering
  \includegraphics[width=0.485\textwidth]{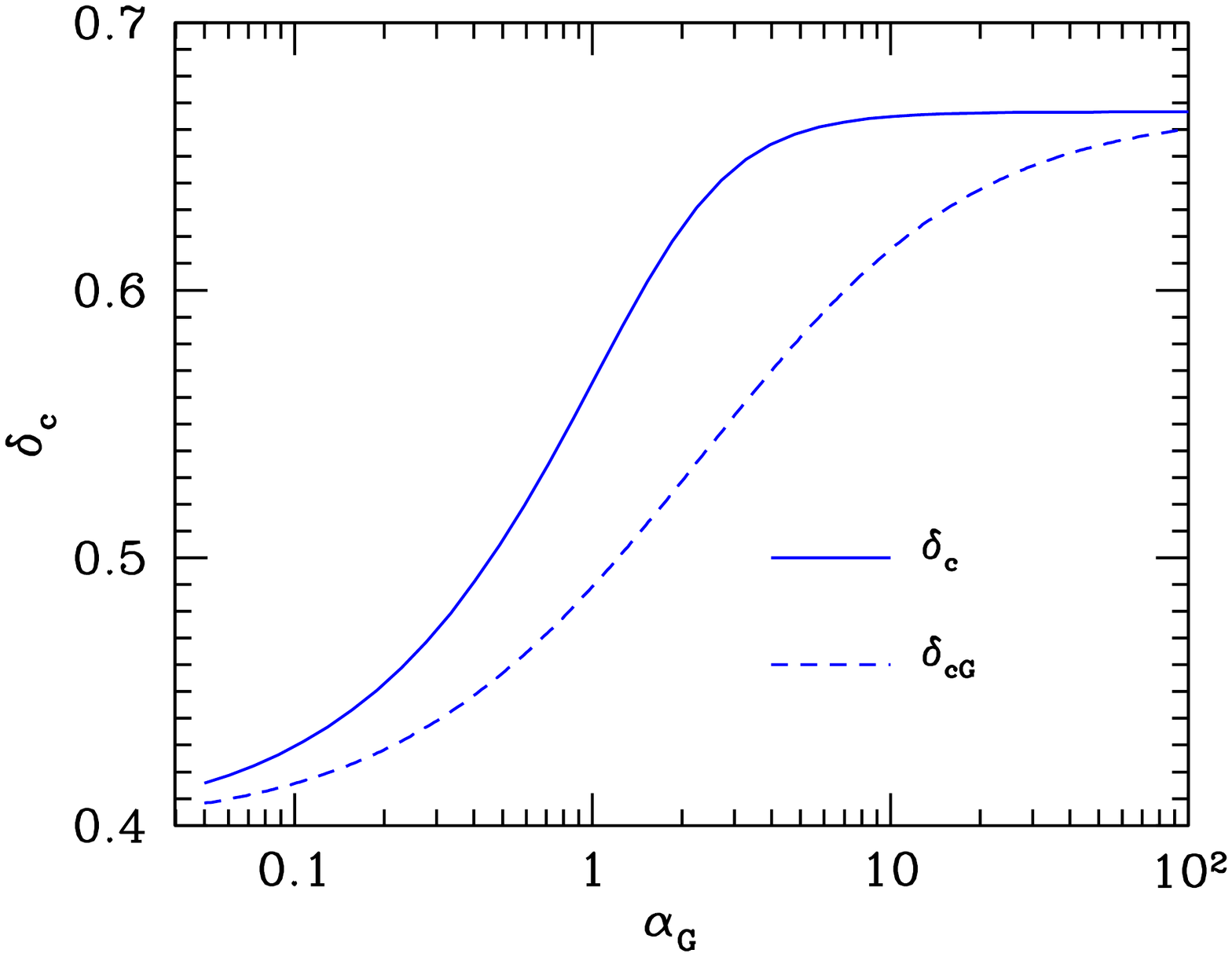} 
  \hspace{0.25cm}
  \includegraphics[width=0.485\textwidth]{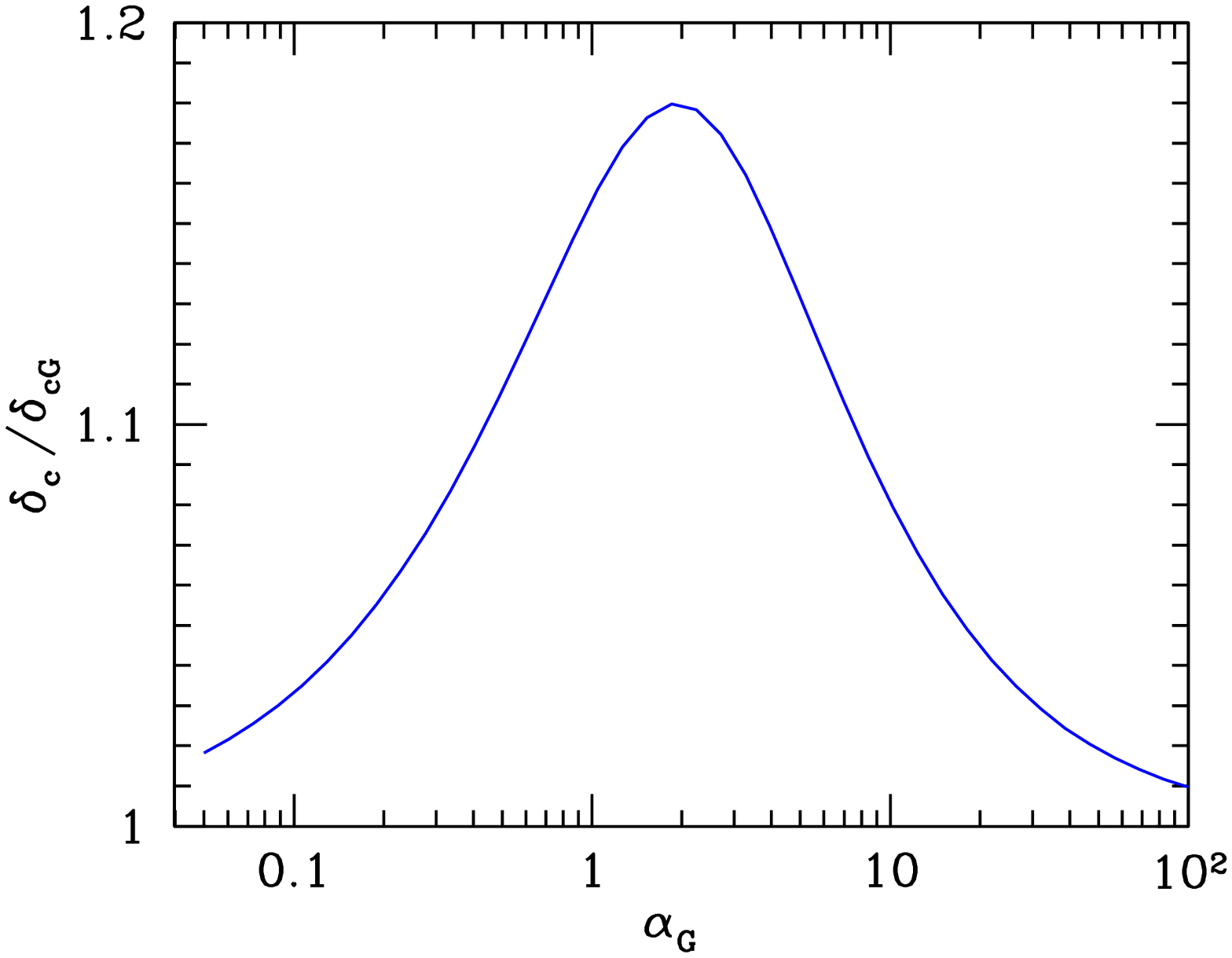} 
\vspace{-4.0cm} 
  \caption{ The left panel of this figure shows the behavior of $\delta_c$ and  $\delta_{c\G} = \delta_c(\alpha_\G)$ 
  plotted as a function of the Gaussian shape parameter $\alpha_\text{\tiny G}$. The right panel shows the ratio of these 
  two quantities: the difference is due to the non linear effects coming from the solution of~\eqref{alpha_sigma}. }
  \label{delta_c2}
 \end{figure*} 

\subsection*{The non linear component of the shape}
If $\zeta$ is a Gaussian random variable, also $\Phi_m$, and  $\Phi_m^{\p\p}\hr_m^2$ obey Gaussian statistics. In such 
case we can write the shape parameter given by~\eqref{alpha_zeta} as
\beq \label{alphaNL}
\alpha = \frac{\alpha_{\text{\tiny G}}}{\left(1 - \frac{1}{2}\Phi_m \right) \left(1-\Phi_m\right)},
\eeq
where
\beq \label{alphaG}
\alpha_{\text{\tiny G}} = - \frac{ \Phi_m^{\prime\prime} \,\hr_m^2 }{ 4\Phi_m} 
\eeq 
is the Gaussian shape parameter obtained in the linear approximation ($\Phi_m \ll 1$), independent of the amplitude 
of $\Phi_m$ since $\Phi_m^{\p\p} \propto \Phi_m$ as, for instance, one can understand by computing the average of  
$\Phi_m^{\p\p}$ given a realisation of $\Phi_m$ using conditional probability. 

The value of  $\Phi_m$ introduces a correction, which is negligible in the linear regime when \mbox{$\Phi_m\ll1$}. On the 
other hand, when the value of $\Phi_m$ is non linear, the term $(1-\Phi_m)(1-\Phi_m/2)$ gives a non negligible modification 
of the value of $\alpha$ with respect $\alpha_\G$. In general $\alpha$ depends on the statistics of $\Phi_m^{\prime\prime}$ 
and the amplitude $\Phi_m$. 

Considering Type I solutions one can write $\Phi_m$ as a function of $\delta_c$ using~\eqref{Phi}, which gives 
\beq
\Phi_m = 1 - \sqrt{1-\frac{3}{2}\delta_c},
\eeq
and then inserting this equation combined with~\eqref{delta_c} into~\eqref{alphaNL} one obtains 
\beq \label{alpha_sigma}
F(\alpha) \left[ 1 + F(\alpha) \right] \alpha = 2 \alpha_{\text{\tiny G}} \,, 
\eeq
where
\beq \label{F_alpha}
F(\alpha) = \sqrt{ 1 - \frac{2}{5} e^{-\sfrac{1}{\alpha}} \frac{\alpha^{1-\sfrac{5}{2\alpha}}}{\Gamma\left(\frac{5}{2\alpha}\right) - 
\Gamma\left(\frac{5}{2\alpha},\frac{1}{\alpha}\right)} } \,.
\eeq
The numerical solution of equation~\eqref{alpha_sigma} gives a value of $\alpha$ as a function of $\alpha_\text{\tiny G}$. 
By inserting this into~\eqref{delta_c}, one can compute the value of $\delta_c$ as a function of $\alpha_{\G}$, which is 
plotted in the left panel of Figure~\ref{delta_c2} using a solid line. This is compared with the analytic behavior of 
$\delta_{c\G} = \delta_c (\alpha_{\G})$ plotted with the dashed line.  

The right panel of Figure~\ref{delta_c2} shows the ratio of these two quantities as function of $\alpha_\text{\tiny G}$, and 
one can appreciate the correction of $\delta_c$ due to the modification of the shape with respect to the one obtained in the 
Gaussian approximation, because of the non linear effects. Because at the boundaries $F(0) \to1$ $(F(\infty) =1)$, there is 
no modification with respect to the Gaussian case and \mbox{$\delta_c = \delta_{c\G}$ in the limits $\alpha \to 0$ 
$(\alpha \to\infty)$}.


\section{The average value of $\delta_c$}
\label{average}
\noindent
The aim of this section is to describe how to calculate the average value of the shape parameter $\alpha$, identifying 
which is the typical perturbation shape associated with a given cosmological power spectrum, which gives the 
corresponding averaged value of the threshold $\delta_c$. 

Assuming Gaussian statistics for the comoving curvature perturbation $\zeta$, the first step is to compute the value 
of $\alpha_\G$ from the power spectrum $P_\zeta(k,\eta)$ defined as 
\begin{equation} \label{Pzeta}
	P_\zeta(k,\eta) = \frac{2 \pi^2 }{k^3}\mathcal{P}_\zeta (k)  T^2(k,\eta), 
\end{equation}	
computed at the proper time $\eta$ when $\hr_m \gg r_H $,  where $r_H = 1 /aH$ is the comoving Hubble radius.
$\mathcal{P}_\zeta(k)$ is the dimensionless form of the power spectrum, and the linear transfer function $T(k,\eta)$,
given by
\begin{equation}
T(k, \eta)= 3 \frac{\sin (k \eta/\sqrt{3})-  (k \eta/\sqrt{3}) \cos (k \eta/\sqrt{3})}{ (k \eta/\sqrt{3})^3} \,,
\end{equation} 
has the effect of smoothing out the subhorizon modes, playing the role of the pressure gradients during the collapse. This
smoothing should be done when $\hr_m \simeq 10\, r_H$ or larger, according to the gradient expansion approach used to 
specify the initial condition of the numerical simulations. This ensures that modes collapsing within the scale $r_H$ does 
not affect the collapse on the larger scale $\hr_m$. The details of how to apply the smoothing have been extensively 
discussed in~\cite{Kalaja:2019uju}, showing that using just the transfer function on superhorizon scales avoids the 
need of introducing a window function on the scale $\hr_m$ of the perturbation, which introduces corrections  in the 
calculation  of the threshold that, however, are reduced when computing the PBH abundance if the same window function is adopted for  evaluating  the variance~\cite{Young:2019osy}.

The radius $\hr_m$ is obtained from condition \eqref{rm_condition}, which can be expressed in terms of the power 
spectrum using Gaussian peak theory to write $\zeta(\hr)$
\beq \label{zeta}
\zeta(\hr) = \zeta_0 \int dk k^2 \frac{\sin(k\hr)}{k\hr} P_\zeta (k,\eta) \,,
\eeq
and, applying $\Phi^\p(\hr_m)=0$, one finally gets
\beq \label{rm_cond}
\int dk k^2 \left[ ( k^2\hr_m^2 - 1 ) \frac{\sin(k\hr_m)}{k\hr_m}  + \cos{(k\hr_m)} \right] P_\zeta(k,\eta) = 0 \,,
\eeq
where this integral equation, in general, has to be solved numerically given the expression of $P_\zeta$.

The Gaussian shape parameter can be computed from the average profile of $\zeta(\hr)$ 
shown in \eqref{zeta}, which allows $\alpha_\G$ to be written as 
\beq
\alpha_{\text{\tiny G}} =  \frac{1}{2} - \frac{\hr_m^2}{4} \frac{\zeta^{\p\p\p}(\hr_m)}{\zeta^\p(\hr_m)},
\eeq
where  we have used the constraint relation $\Phi^\p(\hr_m) = 0$, which gives 
\begin{equation}
 \hr_m^2 \Phi_m^{\p\p} = \hr_m [2\zeta^\p(\hr_m) - \hr_m^2\zeta^{\p\p\p}(\hr_m)] \,.
\end{equation}
Inserting~\eqref{zeta} into the expression for $\alpha_\G$, combined with~\eqref{rm_cond} one obtains
\beq \label{alphaG2}
\alpha_\G = - \frac{1}{4} \left[ 1 + \hr_m \frac{ \int dk k^4 \cos{(k\hr_m)} P_\zeta (k,\eta) }{ \int dk k^3 \sin{(k\hr_m)} 
P_\zeta (k,\eta) } \right], 
\eeq
showing that $\alpha_\G$, and the corresponding value of $\alpha$ computed using~\eqref{alpha_sigma}, are varying
with the shape of the cosmological power spectrum. The same holds for the value of $\hr_m$ given by the solution 
of~\eqref{rm_cond}. The values of $\alpha_\G$ and $\alpha$ can then be used in~\eqref{delta_c} so as to calculate the 
corresponding values of $\delta_{c\G}$ and $\delta_c$, obtaining a direct relation between the threshold 
and the particular shape of the cosmological power spectrum $P_\zeta$. 

In the following we are going to apply this prescription to study the extent to which, given a particular form of the power 
spectrum, the amplitude of the threshold $\delta_c$ is varying.

 \begin{figure*}[ht!]
 \vspace{-1cm}
 	\centering
 	\includegraphics[width=0.485\textwidth]{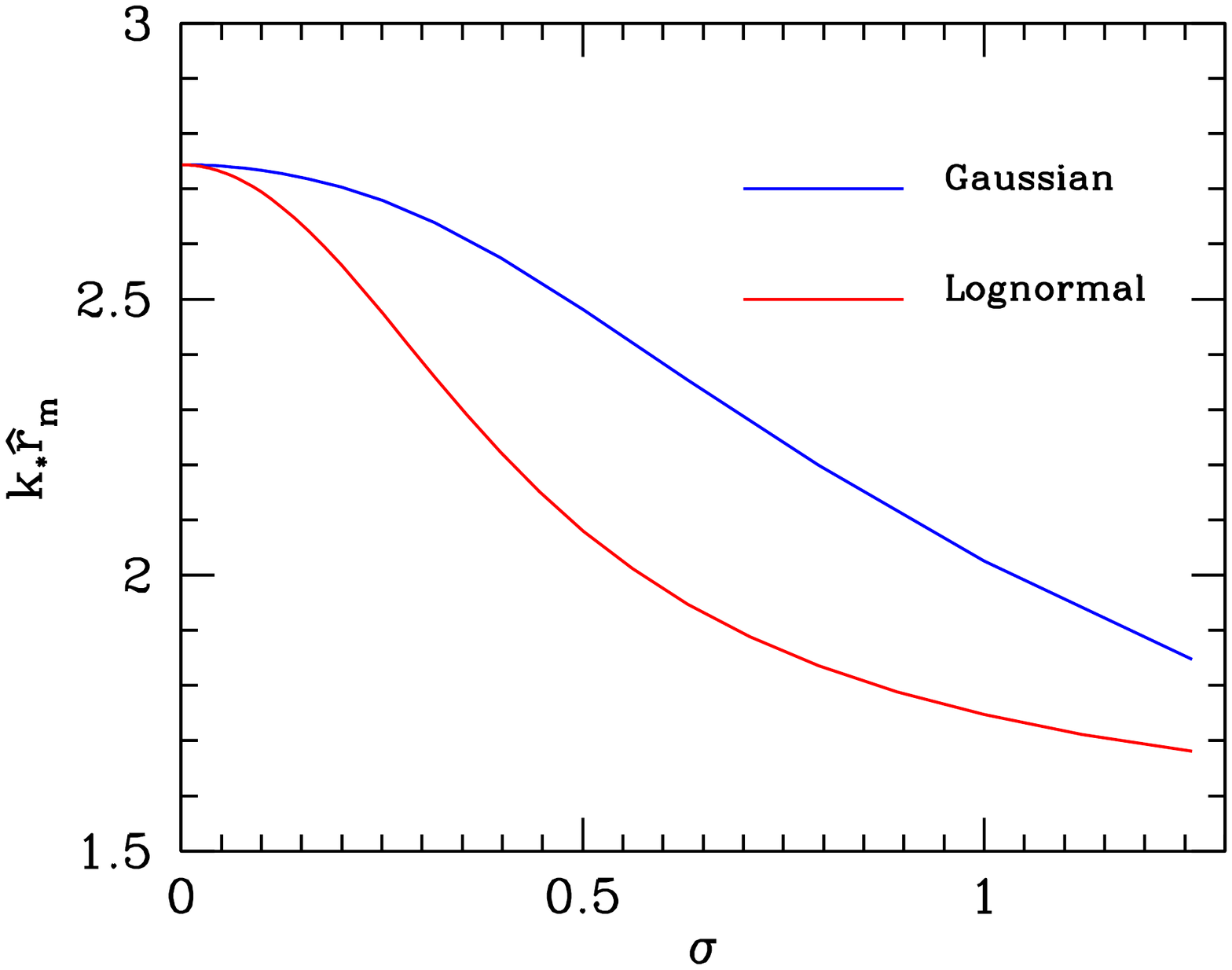} 
	\hspace{0.25cm}
 	\includegraphics[width=0.485\textwidth]{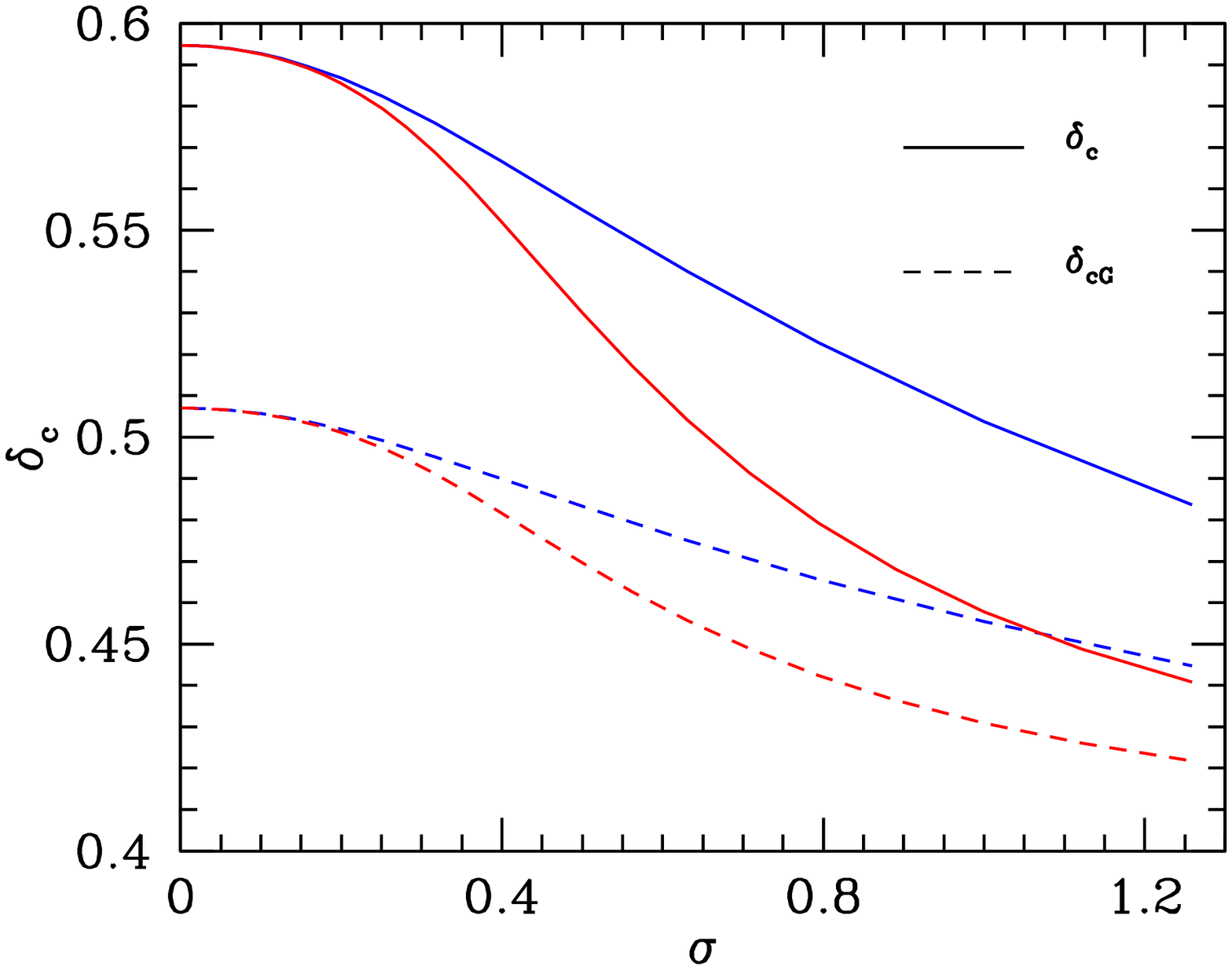} 
	\vspace{-4.0cm}
 	\caption{ {\it Left:} Peak position  of the compaction function $\hr_m$ for both the case of a gaussian and lognormal 
 		shape of the power spectrum. 
 		{\it Right:} Average threshold for collapse for both the $\delta_{c\text{\tiny G}}(\alpha_\text{\tiny G})$ and its 
 		corresponding non-linearly corrected $\delta_c$ in both cases.
 		In the limit of $\sigma \to 0$, both spectra converge to the monochromatic case and the result is compatible 
 		with the  Dirac delta example. }
 	\label{Lognormal}
 \end{figure*}

\subsection*{Peaked Power Spectrum}
The simplest cosmological power spectrum of the comoving curvature perturbation  that can be considered is monochromatic,  
behaving like a Dirac-delta distribution, typically written as 
\be
\label{dirac}
{\cal P}_\zeta (k)={\cal P}_0 k_*\delta_D(k-k_*) \,.
\ee
Inserting this into~\eqref{rm_cond} we get $k_* \hr_m \simeq 2.74$, which gives $\delta_{c,\text{\tiny G}} \simeq 0.51$, a value 
of the threshold in the Gaussian approximation consistent with the one obtained in~\cite{Germani:2018jgr}.  Solving 
equation~\eqref{alpha_sigma}, we can see the corresponding modification of $\delta_c$ due to the non linear effects, giving 
$\alpha \simeq  6.33$ corresponding to $\delta_c \simeq 0.59$. 

It is interesting to note that this value of the threshold is consistent with the one found if the average profile of $\zeta(r)$ 
for a peaked power spectrum, characterised by the sync function, is inserted into~\eqref{rel} to specify the initial conditions 
for the numerical simulations \cite{Kehagias:2019eil}. This is consistent with the fact that using peak theory in $\zeta$ or in 
the density contrast $\delta\rho/\rho_b$ is equivalent when the power spectrum is very peaked, behaving like a Dirac 
delta~\cite{DeLuca:2019qsy}.

\subsection*{Broad Power Spectrum} 
A class of models with a broad and flat power spectrum of the curvature perturbations of the 
form~\cite{MoradinezhadDizgah:2019wjf, DeLuca:2020ioi}
\be
\label{broad}
{\cal P}_\zeta (k)={{\cal P}_0} \Theta (k - k_\text{\tiny min}) \Theta (k_\text{\tiny max} - k), \quad k_\text{\tiny max} \gg k_\text{\tiny min}
\ee
is another simple toy model, corresponding to the top hat shape of the  primordial power spectrum, which is considered 
in~\cite{Germani:2018jgr}.  In this case, from~\eqref{rm_cond} we have $k_\text{\tiny max}\hr_m\simeq 4.49$, which gives 
$\alpha_{\text{\tiny G}} \simeq 0.9$ and $\delta_{c,\text{\tiny G}} \simeq  0.48$. 

The values of $k_\text{\tiny max}\hr_m$ and $\delta_{c,\G}$ obtained here are different from the values 
$k_\text{\tiny max}\hr_m\simeq 3.5$ and $\delta_{c,\text{\tiny G}} \simeq  0.51$ found in~\cite{Germani:2018jgr}, because
in that analysis peak theory was applied directly to the linearised density contrast $\delta\rho/\rho_b$ while here, instead, 
we are using peak theory to compute the average curvature perturbation $\zeta$ and account for the non-linear relation 
with the compaction function. For this reason the integrals in peak theory for finding $\hr_m$ and the shape profile 
$\zeta(\hr)$ are characterised by a higher power in $k$. 

Solving equation~\eqref{alpha_sigma} to include the non linear effects gives $\alpha \simeq 3.14$ corresponding to 
$\delta_c \simeq 0.56$.

\subsection*{Gaussian Power Spectrum} 
The gaussian shape of the curvature power spectrum given by
\bea
{\cal P}_\zeta (k) & = &{\cal P}_0  \exp{ \left[ - (k-k_*)^2 /2\sigma^2 \right] },
\eea
is characterised by the central reference scale $k_*$ and width $\sigma$. Solving \eqref{rm_cond}, the relation between the 
length scale $\hr_m$ of the perturbation and the scale $k_*$ is shown in the left panel of  Fig.~\ref{Lognormal}. As one can
appreciate, in the limit of the narrow case $\sigma \to 0$ the result converges to the one obtained for a monochromatic 
shape of the curvature power spectrum (studied previously in the peaked case), while for broader shapes the expected length 
scale of the overdensity multiplied by $k_*$ is decreasing. This is a result of the fact that, for broader shapes, more modes are 
contributing to the collapse, resulting in a narrower curvature profile. 

The behavior of the shape parameter $\alpha$, which decreases as $\sigma$ increases, reflects the fact that when multiple 
modes are participating in the collapse, the compaction function becomes flatter. As a consequence, the pressure gradients 
are reduced, facilitating the collapse, and the corresponding threshold for PBHs decreases for larger values of $\sigma$, 
as one can appreciate in the right panel of the same figure. As discussed in the previous section and shown in Fig.~\ref{delta_c2}, 
as non-linearities are taken into account, the critical threshold $\delta_c$ reaches larger values than tat for $\delta_{c\G}$
computed in the Gaussian approximation.

\subsection*{Lognormal Power Spectrum} 
The lognormal power spectrum is expressed as
\bea
{\cal P}_\zeta  (k) & = & {\cal P}_0  \exp{ \left[ - \ln^2{(k/k_*)} /2\sigma^2 \right]},
\eea
characterised by a width $\sigma$ and a central scale $k_*$. The relation between the length scale of the overdensity 
and the scale $k_*$ is plotted in the left panel of Fig.~\ref{Lognormal}, while the right panel is showing the average threshold 
for PBHs, showing the same qualitative behaviors found for the Gaussian power spectrum.

Because $\sigma$ in this case identifies the width of the power spectrum in logarithmic space, larger values of $\sigma$ allow 
for more modes to be part of the collapse. As a consequence, if compared to the Gaussian case, the trends for the relative 
change of $k_* \hr_m$ and $\delta_c$ are amplified.

\begin{figure*}[ht!]
	\centering
	\vspace{-1cm}
	\includegraphics[width=0.485\textwidth]{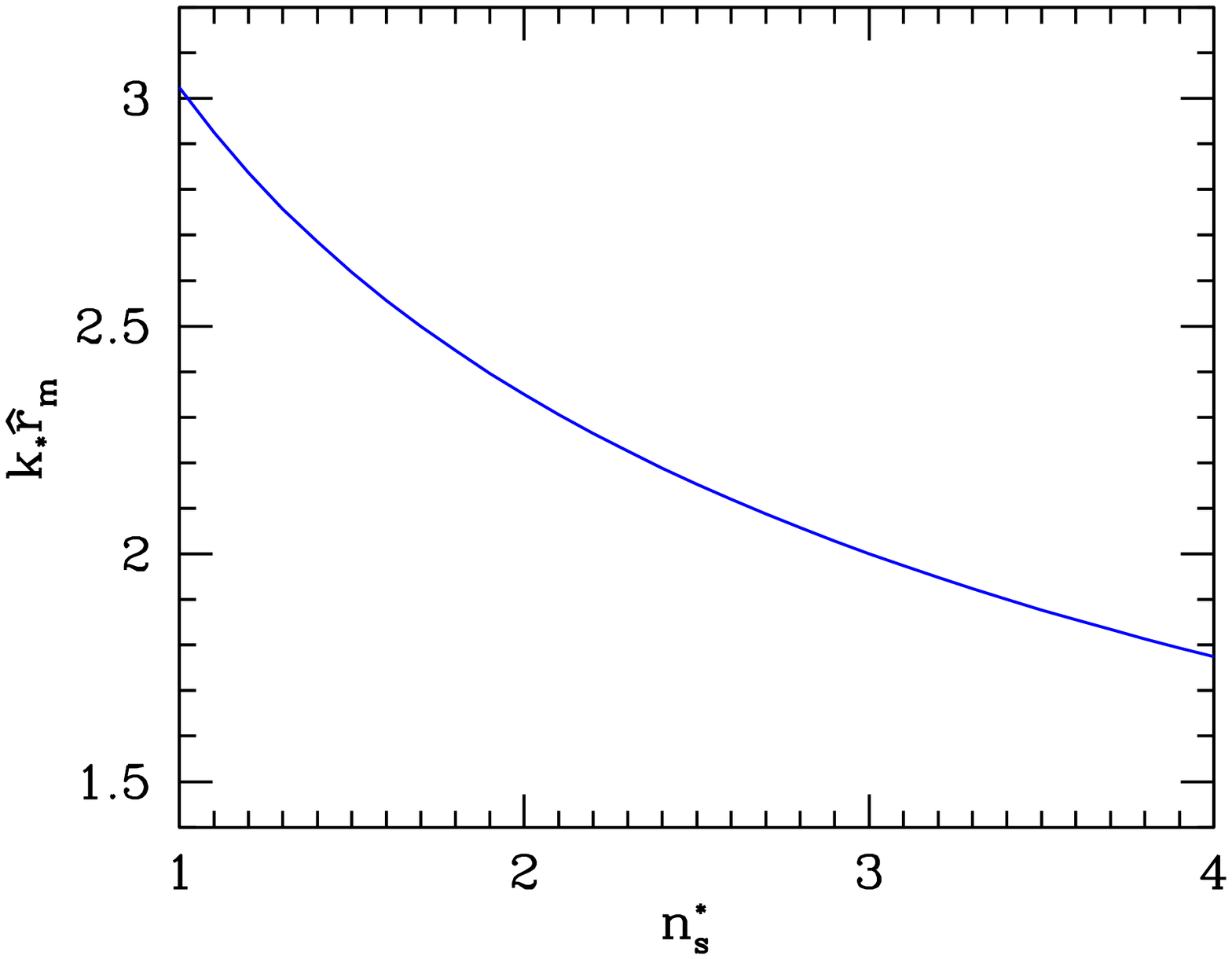} 
	\hspace{0.25cm}
	\includegraphics[width=0.485\textwidth]{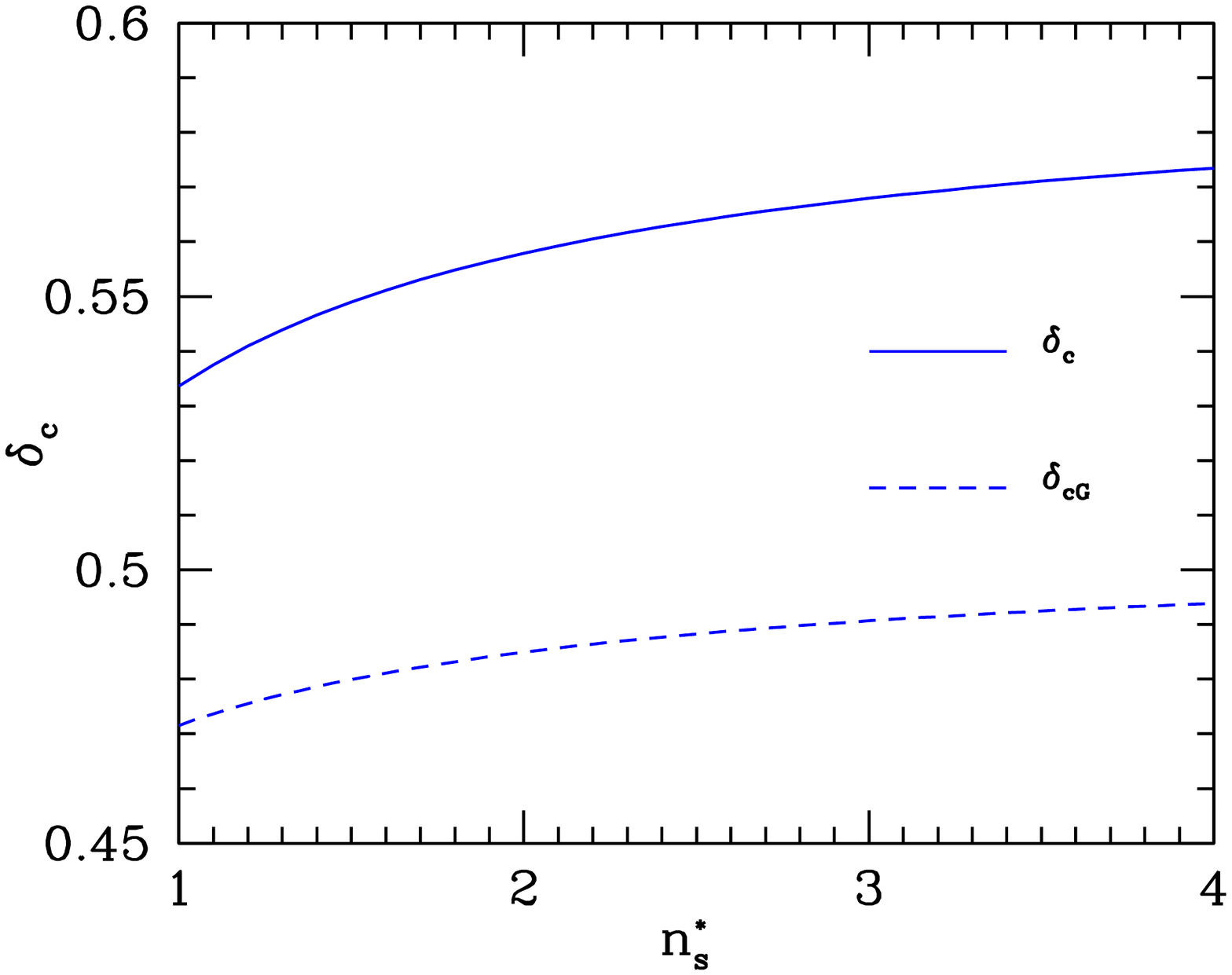} 
	\vspace{-4cm}
	\caption{The same as in Fig.~\ref{Lognormal}, but for the cut-power-law power spectrum.}
	\label{cutpowerlaw}
\end{figure*}

\subsection*{Cut-Power-Law Power Spectrum} 
The cut-power-law curvature power spectrum is given by
\bea
{\cal P}_\zeta (k) & = &  {\cal P}_0 \left(\frac{k}{k_*}\right)^{n_s^*} \exp{\left[- (k/k_*)^2 \right]},
\eea
expressed in terms of a tilt $n_s^*$ and with an exponential cut-off at the momentum scale $k_*$.
The relation between the length scale of the overdensity and the scale $k_*$ is shown in the left panel of Fig.~\ref{cutpowerlaw}, 
while the right panel of this figure is showing the behavior of the average threshold for PBHs.  

As $n_s^*$ increases, the spectrum becomes narrower, with a shift towards  a higher value of the power spectrum peak which 
is identified by the maximum of the combined product of $k^{n_s^*}$ and the exponential cut-off. In agreement with the behavior 
seen in the previous examples, as the spectral tilt decreases, a larger number of modes participate in the collapse, resulting in a 
lower value of the threshold $\delta_c$.

\subsection*{Summary} 
The analysis of this section of different power spectra shows that, when the shape is broader, the value of the threshold 
$\delta_c$ is lower because more modes are involved in the collapse. The maximum value we have found is $\delta_c \simeq 0.59$ 
when the power spectrum behave like a Dirac Delta (corresponding to a single mode).
The behaviour for the lognormal power spectrum that one can extrapolate looking at the right panel of 
Figure~\ref{Lognormal} indicates the possibility of getting closer to the lower boundary of $0.4$ for very large values of $\sigma$. 
In conclusion the shapes of the power spectra we have considered here allows $ 0.4 \lesssim \delta_c \lesssim 0.6$, when the 
threshold is computed on superhorizon scales.

\begin{figure*}[t!]
	\vspace{- 1.0cm}
	\centering
	\includegraphics[width=0.48\textwidth]{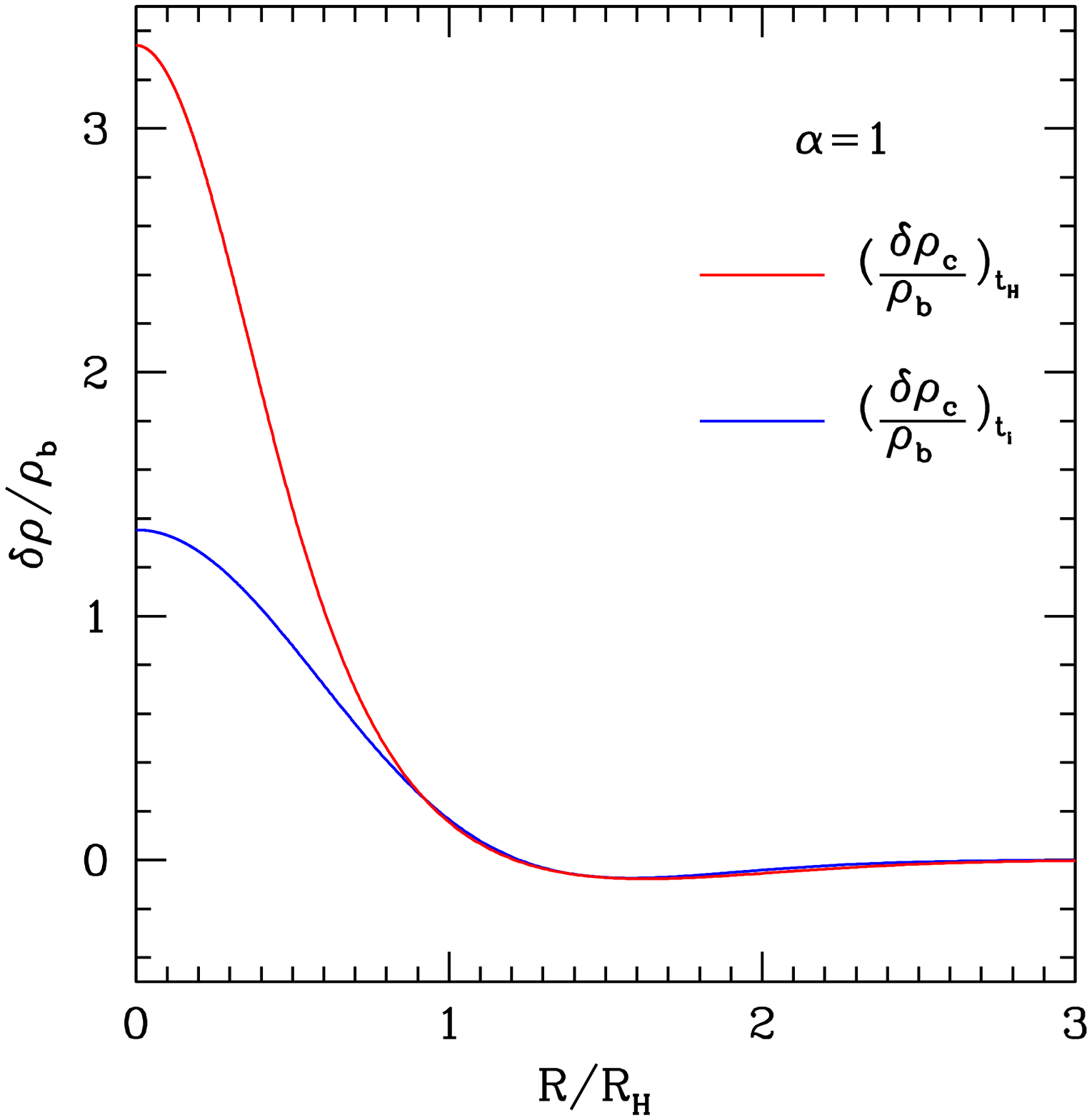} \hspace{0.4cm}
	\includegraphics[width=0.48\textwidth]{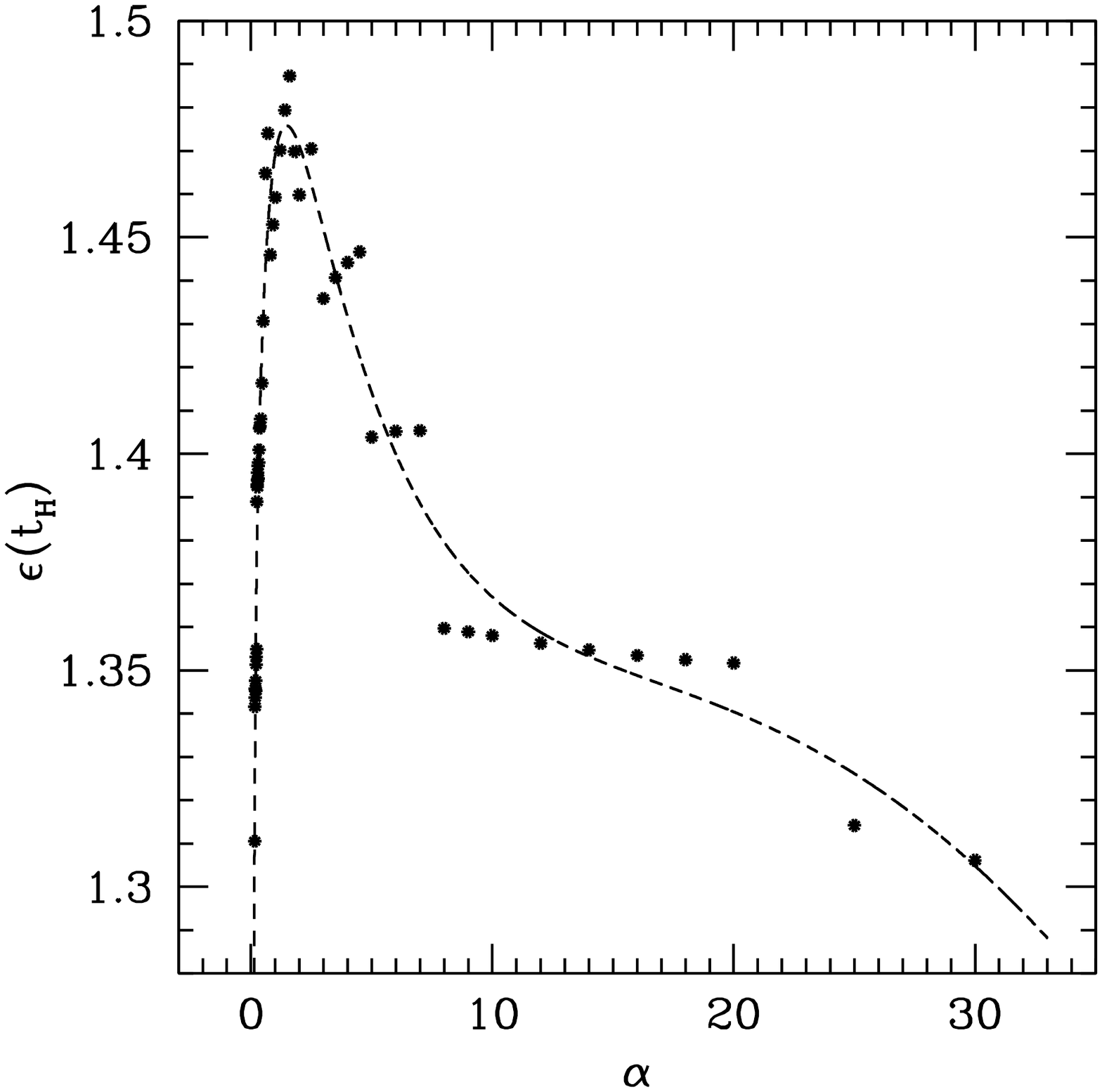} 
	\vspace{-2.5cm}
	\caption{ The left panel shows the critical Mexican-hat profile of the energy density, obtained from~\eqref{K_basis} 
		with $\alpha=1$, computed at the horizon crossing, linearly extrapolated ($\epsilon=1$) with a blue line, and  
		computed numerically at the non linear horizon crossing (red line), corresponding in this case to 
		$\epsilon\simeq1.46$. Both profiles are plotted against $R/R_H$, where $R_H$ is the radius of the 
		cosmological horizon computed at the corresponding time. The right panel shows how the non linear horizon 
		crossing, measured in terms of $\epsilon$, varies when plotted against the shape parameter $\alpha$, compared 
		to $\epsilon=1$ at the linear horizon crossing. The dashed line is a polynomial fit of numerical data given by 
		the dots. } 
	\label{rho1}
\end{figure*}


\section{The non linear horizon crossing}
\label{non_linear_horizoncrossing}
In this section we study the effects on the threshold when the cosmological horizon crossing is computed during the 
numerical evolution, measuring the amplitude $\delta_m$ of the perturbation when the length scale $R_m$ is equal to 
the cosmological horizon radius $R_H$ defined with respect to the perturbed medium. The numerical code used for 
the simulations is the same as used in previous works (see \cite{musco} and references therein for more details).

The threshold for PBHs has so far been computed at cosmological horizon crossing by making a linear extrapolation from 
the superhorizon regime, where the curvature is time independent, imposing $aHr_m=1$ in equation~\eqref{rel}, where 
the cosmological horizon $R_H = 1/H$ is defined with respect to the background. In this way one is extending the validity 
of the gradient expansion approximation up to $\epsilon=1$, which is not very accurate. Although this represents a well 
defined criterion for measuring the perturbation amplitude, and has been widely used in the literature to compute the threshold 
$\delta_c$ for PBHs, it does not give the correct amplitude of the perturbation at the ``real" cosmological horizon crossing, 
because it is neglecting the non linear effects of the higher orders in the gradient expansion approach, which are taking 
into account that the curvature profile $\zeta$ starts to vary with time when $\epsilon \sim 1$. 

In general the cosmological horizon is a marginally trapped surface within an expanding region, which in spherical symmetry 
is simply defined by the condition $R(r,t)=2M(r,t)$, where $R(r,t)$ is the areal radius and $M(r,t)$ is the mass within a given 
sphere of radius $R(r,t)$, called the Misner-Sharp mass. This relation for a trapped surface is very general, assuming only 
spherical symmetry, and allows computation of the location of any apparent horizon: if we have an expanding medium, this 
is a cosmological horizon, while if the medium is collapsing then it is a black hole apparent 
horizon~\cite{Helou:2016xyu,Faraoni:2016xgy}. 

In simulations of PBH formation, because we are in a locally closed Universe, the rate of expansion of the cosmological 
horizon is less than that of the spatially flat background, and this gives rise to an additional growth of the amplitude of the
perturbation before reaching the horizon crossing. 

The left plot of Figure~\ref{rho1} shows the critical energy density profile obtained with the curvature perturbation given 
by~\eqref{K_basis}, with $\alpha=1$ corresponding to a Mexican-hat shape, computed at the horizon crossing linearly 
extrapolated (blue line) and at the non linear horizon crossing obtained from the numerical simulations (red line). The 
second profile shows an additional growth of the amplitude, which is not negligible when the value of the energy density 
obtained with the linear extrapolation is non-linear. Part of this extra growth is due to the longer time necessary to reach 
the non linear horizon crossing which can be seen explicitly in the right plot where $\epsilon(t_H)$ is plotted against the 
shape parameter $\alpha$, with the dashed line fitting the numerical results given by the dots. We can appreciate that the 
value of $\epsilon(t_H)$ at the non linear horizon crossing, $1.3 \lesssim \epsilon(t_H) \lesssim 1.5 $, is larger than one 
given by the linear horizon crossing ($aHr_m=1$). In particular, for $\alpha=1$ the non linear horizon crossing is obtained 
at $\epsilon(t_H)\simeq1.46$, corresponding to an amplitude of the central peak calculated with \eqref{rel} equal to 
$\delta\rho_0/\rho_b\simeq1.98$, as compared with the value of $\delta\rho_0/\rho_b\simeq1.35$ computed at the linear 
horizon crossing (blue line). The additional growth of the profile, with a peak value of the density contrast 
$\delta\rho_0/\rho_b\simeq3.34$ obtained numerically at the non linear horizon crossing (red line), is explained by the 
higher orders in the gradient expansion which need to be taken into account when $\epsilon \sim 1$. This effect is 
genuinely non linear. 

The delay of the horizon crossing due to the non linear effects gives an increase of the final mass of PBHs because of 
the corresponding increase of the cosmological horizon mass $M_H$: during the radiation dominated universe 
$M_H \sim \epsilon^2$ and from Figure~\ref{rho1} one can appreciate the change of the horizon scale introducing a 
correction value about equal to 2 (i.e. $1.7 \div 2.2$) in the cosmological horizon mass.

\begin{figure*}[t!]
	\vspace{-1.0cm}
	\centering
	\includegraphics[width=0.48\textwidth]{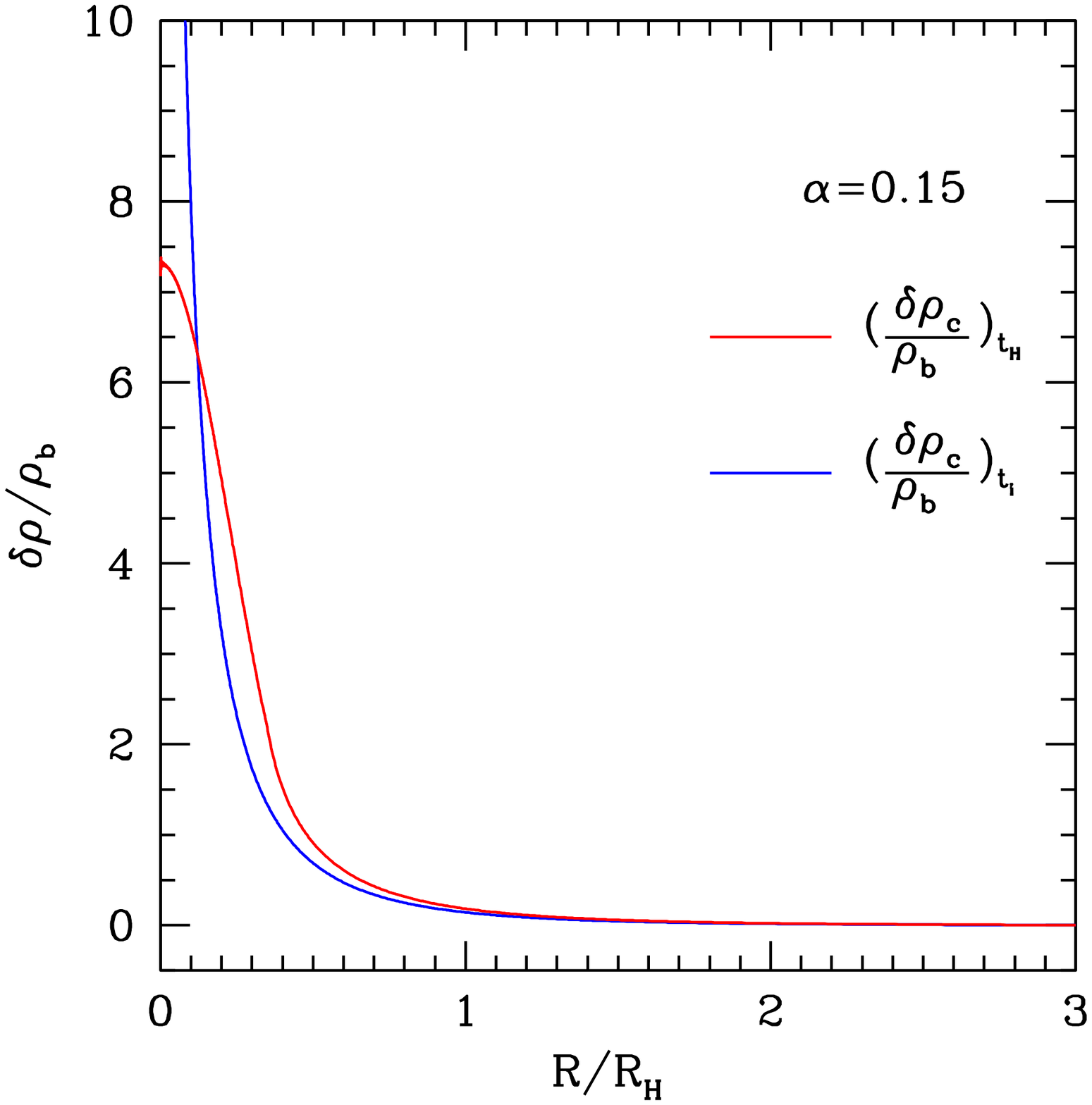} \hspace{0.4cm}
	\includegraphics[width=0.48\textwidth]{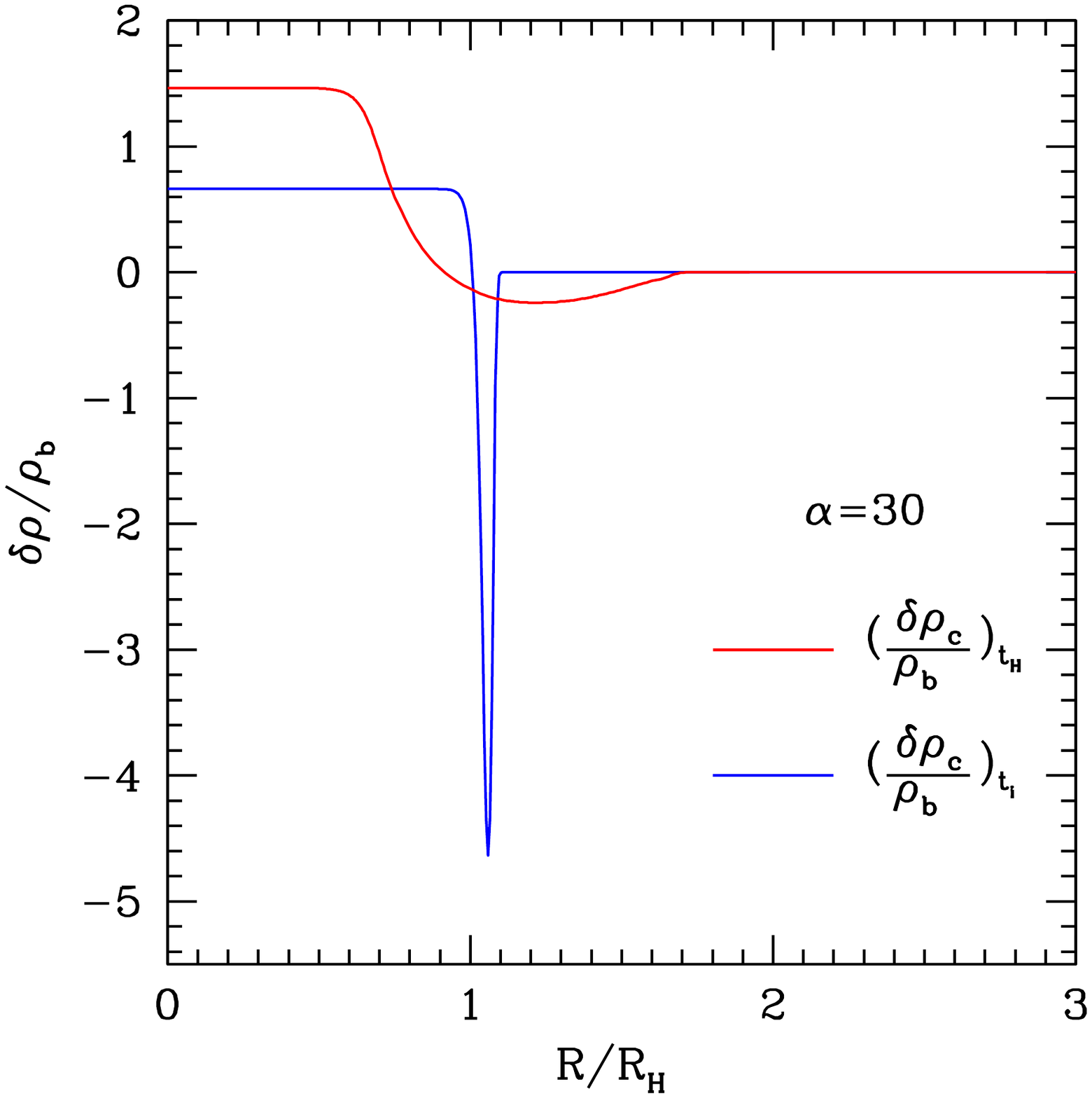} 
	\vspace{-2.5cm}
	\caption{The two plots of this figure show the critical energy density profiles obtained from~\eqref{K_basis} with 
		$\alpha=0.15$ (left panel) and $\alpha=30$ (right panel), plotted against $R/R_H$, computed at the horizon 
		crossing linearly extrapolated (blue line) and at the non linear one (red line). }
	\label{rho_alpha}
\end{figure*}

 In the left panel of Figure~\ref{rho_alpha} we are comparing the critical energy density profiles obtained from~\eqref{K_basis} 
 for \mbox{$\alpha=0.15$}, which gives a very sharp profile, almost like a Dirac-delta, while in the right panel we plot the 
 critical profiles computed for $\alpha=30$, which gives a very broad profile, very similar to a top-hat. As with the Mexican-hat 
 shape, the profile in the right frame computed at the non linear horizon crossing is characterised by an extra growth of the 
 peak: the numerical evolution gives $\delta\rho_0/\rho_b\simeq1.46$ at $\epsilon(t_H)\simeq1.31$ as compared with 
 $\delta\rho_0/\rho_b\simeq0.66$ obtained at $\epsilon=1$ with the linear extrapolation. As in Figure~\ref{rho1} for $\alpha=1$, 
 this difference is a result of the combination of the extra linear growth due to the larger value of $\epsilon$ and the non linear 
 effects.
 
 For the very sharp profile plotted in the left panel ($\alpha=0.15$) we can observe instead that the value of the peak 
 amplitude is significantly reduced at the non-linear horizon crossing with respect the one computed with a linear 
 extrapolation at $\epsilon=1$. This is because for $\alpha=0.15$ the profile is not smooth in the center, and there is 
 a significant effect of the local pressure gradients, which are smoothing the profile during the evolution, giving at the 
 non linear horizon crossing time a smooth profile with a much lower amplitude of the peak: $\delta\rho_0/\rho_b\simeq7$ 
 as compared with $\delta\rho_0/\rho_b\simeq338$ linearly extrapolated at $\epsilon=1$. A similar effect happens in the under 
 dense region for the top-hat like profile ($\alpha=30$). 
 
 In Figures~\ref{rho1} and \ref{rho_alpha} we have analysed three sample cases of the energy density profiles, seeing 
 how the shape is modified at the non linear horizon crossing with respect to the one imposed at initial conditions on super 
 horizon scales, discovering the following general behavior: if the profile is initially smooth, the peak amplitude computed 
 at the non linear horizon crossing is higher than the one extrapolated linearly due to non linear effects which give an extra 
 growth factor, while when the profile is sharp the behavior is the opposite, due to the non linear effects of the pressure 
 gradients smoothing the profile. In general very large values of the peak amplitude at horizon crossing are strongly 
 suppressed because of the smoothing induced by the pressure gradients. 
 
\begin{figure*}[t!]
 \vspace{-1.0cm}
 \centering
  \includegraphics[width=0.48\textwidth]{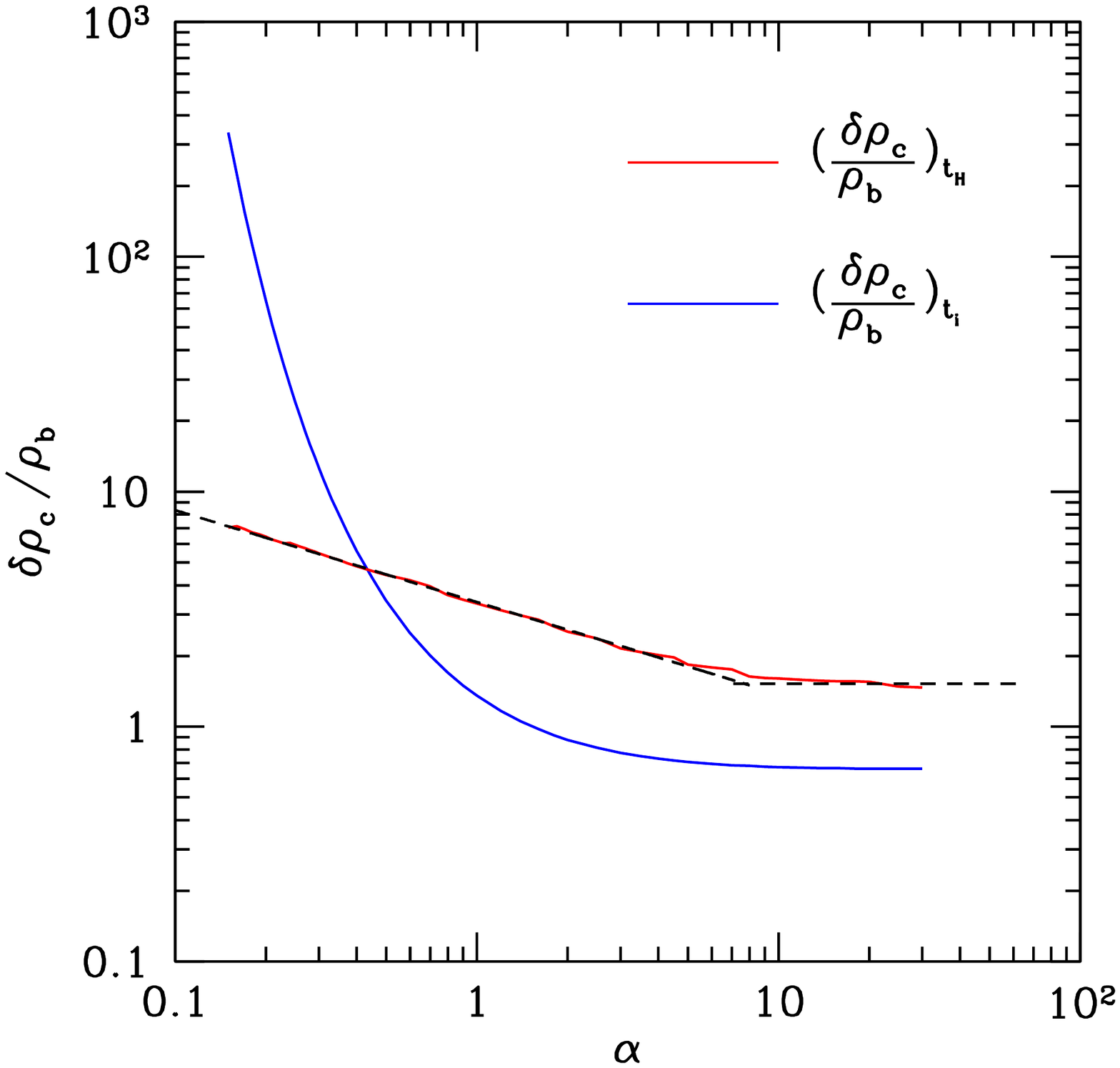} \hspace{0.4cm}
  \includegraphics[width=0.48\textwidth]{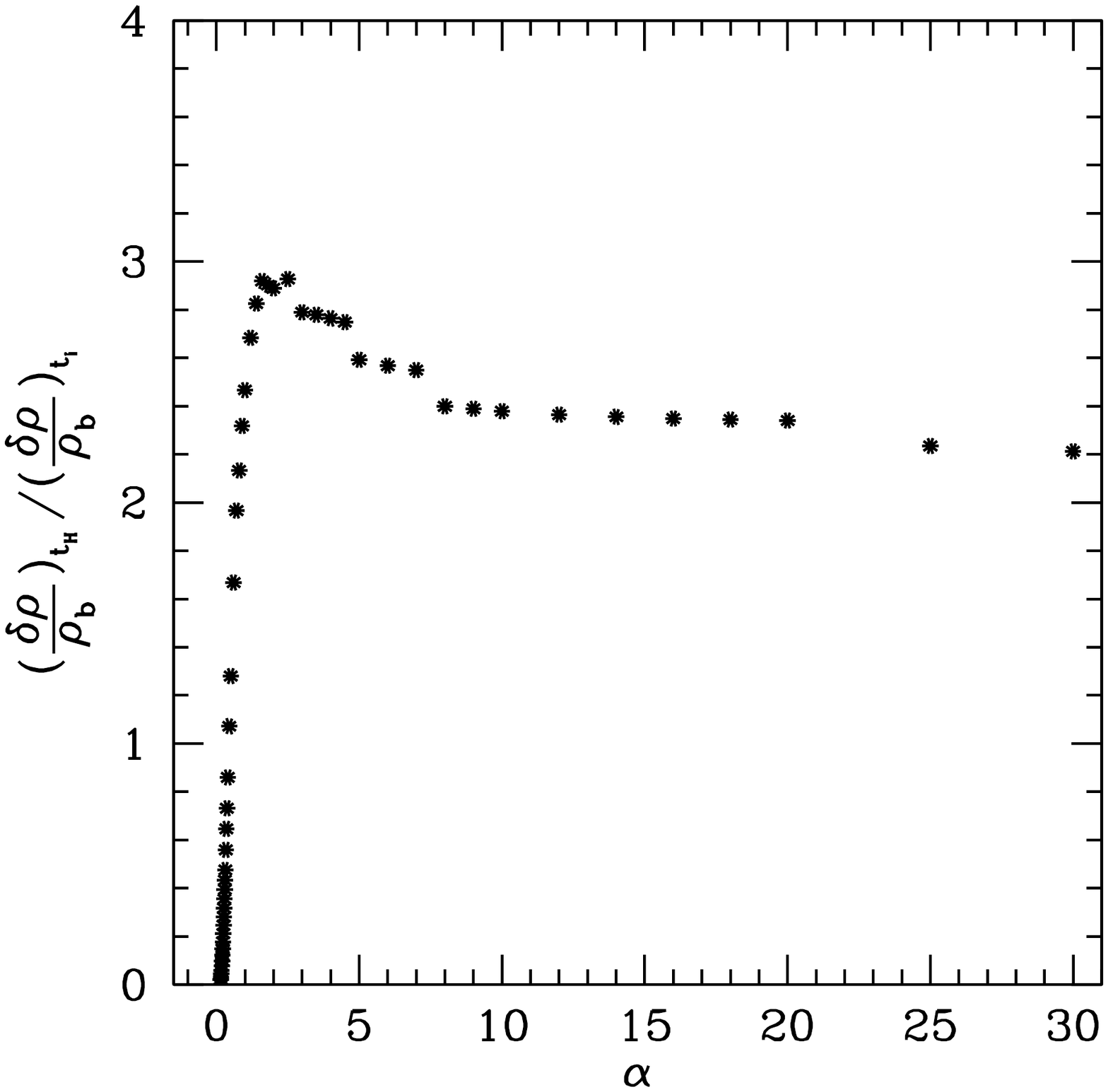} 
  \vspace{-2.5cm}
  \caption{ The left panel of this figure shows the two behaviors of the critical amplitude of the peak $\delta\rho_c/\rho_b$, 
  in one case extrapolated linearly at horizon crossing (blue line) and in the other one computed at the non linear horizon 
  crossing (red line), plotted as function of the shape parameter $\alpha$. The right panel of this figure shows the 
  corresponding ratio of these two quantities.}
  \label{peak_c}
 \end{figure*}

 In general the critical amplitude of the peak $\delta\rho_c/\rho_b$ depends on the shape, and in the left panel of 
 Figure~\ref{peak_c} we can see how this quantity is varying with respect to $\alpha$, for all of the range of shape described 
 by $0.15\leq\alpha\leq30$. The linearly extrapolated  values of the critical amplitude of the peak, given by~\eqref{peak}, are 
 plotted with a blue line, while the values computed at the non linear horizon crossing are plotted with a red line.
 
 The linearly extrapolated critical peak values can be computed analytically from~\eqref{peak} while, as shown in the plot, 
 the critical values computed at the non-linear horizon crossing are given with a good approximation by a simple fit, 
 divided in two regimes.
 \beq \label{delta_rho_fit}
 \frac{\delta\rho_c}{\rho_b} \simeq
 \left\{
\begin{aligned}
& 10^{\,0.53-0.17\ln \alpha}  \quad &\alpha \lesssim 8 \\ 
& 1.52 \quad  &\alpha \gtrsim 8
\end{aligned}
\right.
 \eeq 

In the right panel of Figure~\ref{peak_c} we show the ratio between the critical amplitude computed at the non linear 
horizon crossing and the one linearly extrapolated. This shows clearly the two different regimes: the first one, for
\mbox{$\alpha\lesssim8$}, with the critical amplitude varying with $\alpha$, and the second one for \mbox{$\alpha\gtrsim8$}
which is almost independent of $\alpha$, with the peak amplitude converging towards an almost constant value. 

The linearly extrapolated value is equal to the one computed numerically for $\alpha \simeq 0.45$, because the energy 
density profiles obtained from~\eqref{K_basis} are not smooth if $\alpha \leq 0.5$, with a non vanishing first 
derivative in the center. On the contrary, for $\alpha > 0.5$ the energy density profiles are smooth in the center and the 
perturbation is free to grow without any relevant smoothing of the shape produced by the pressure gradients, reaching a 
larger value of the critical peak amplitude at the non linear horizon crossing with respect to the one linearly extrapolated. 

\begin{figure*}[t!]
 \vspace{-1.0cm}
 \centering
   \includegraphics[width=0.48\textwidth]{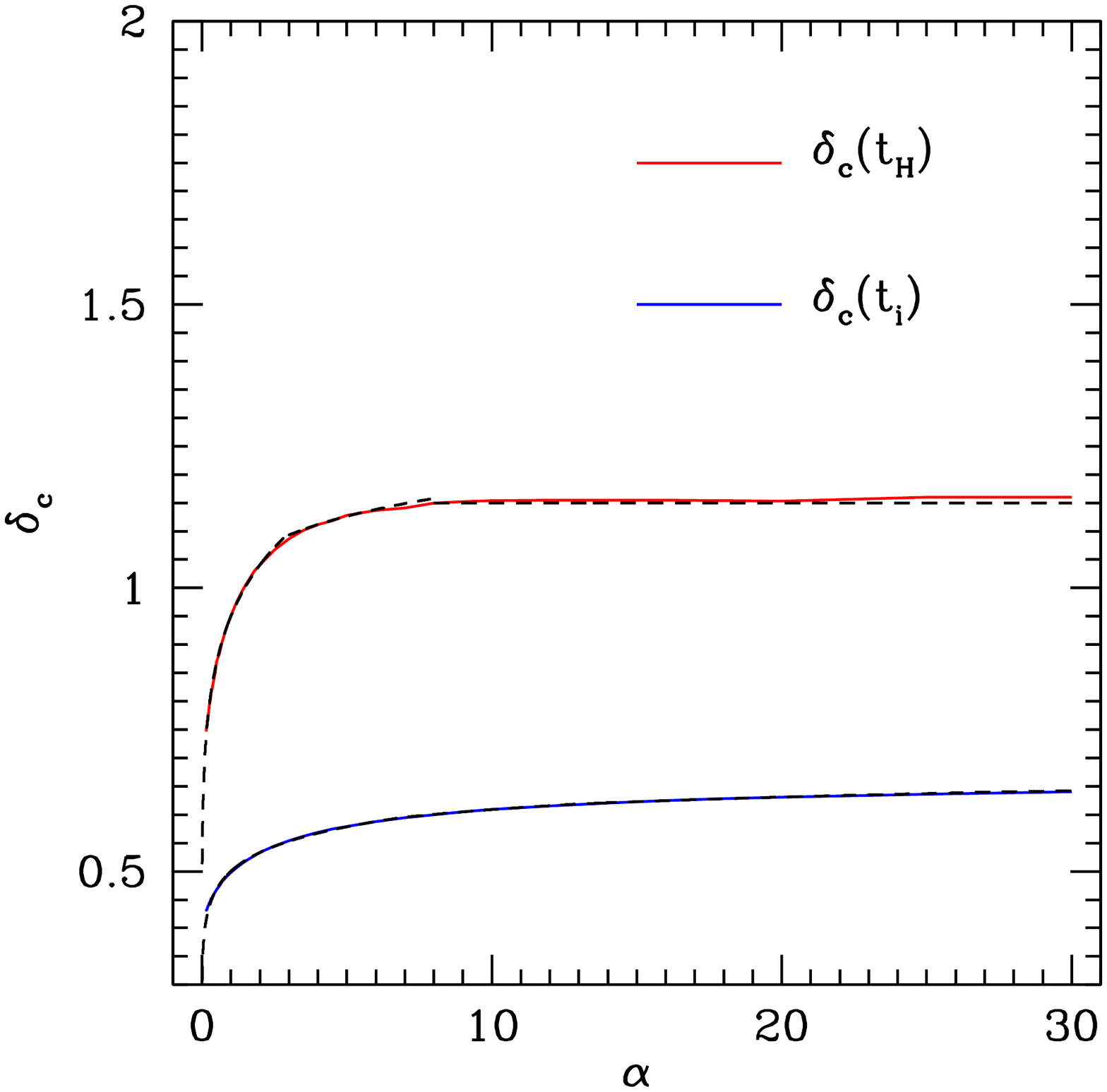} \hspace{0.4cm}
  \includegraphics[width=0.48\textwidth]{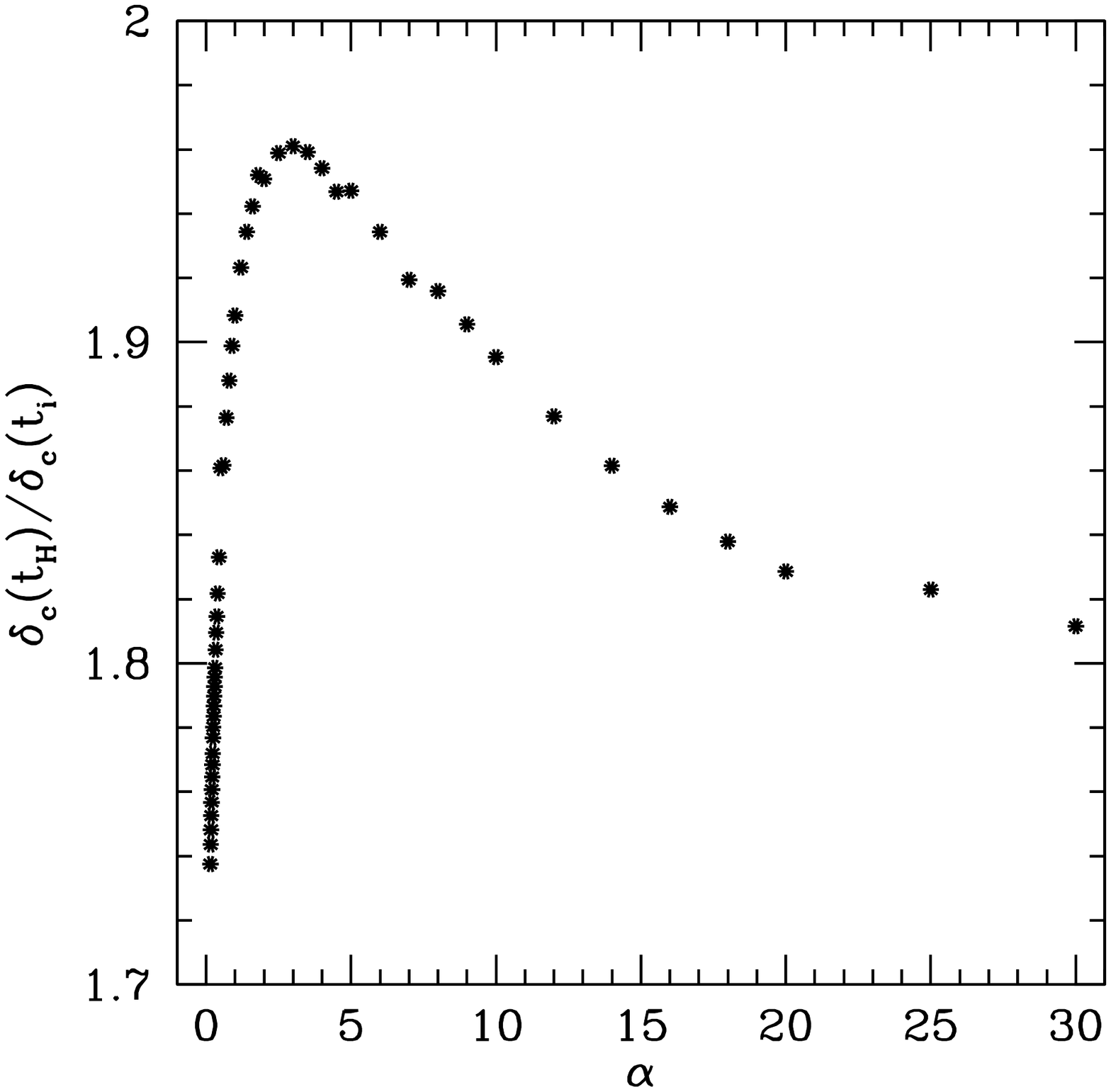} 
  \vspace{-2.5cm}
  \caption{ The left panel of this figure shows the two behaviors of the threshold $\delta_c$ in one case extrapolated 
  linearly at horizon crossing (blue line) and in the other one computed at the non linear horizon crossing (red line), 
 plotted as a function of the shape parameter $\alpha$. The right panel of this figure shows the corresponding ratio of 
 these two quantities.}
  \label{fig:delta_c}
 \end{figure*}

In Figure~\ref{fig:delta_c} the same analysis is made for the threshold $\delta_c$, with the left plot showing the threshold 
$\delta_c(t_i)$ linearly extrapolated (blue line) and the threshold $\delta_c(t_H)$ computed at the non linear horizon 
crossing (red line). The linearly extrapolated threshold, described with a very good approximation by the analytic expression 
of equation \eqref{delta_c}, can be divided into three different regimes, each one described by a simple fit.
\beq 
\label{delta_c_fit}
 \delta_c (t_i) \simeq
 \left\{
\begin{aligned} 
& \alpha^{0.047} - 0.50  \quad &0.1\lesssim \, \alpha \lesssim \ 7 \,  \\ 
& \alpha^{0.035}  - 0.475  \quad\quad  &7\lesssim \, \alpha \lesssim 13  \\ 
& \alpha^{0.026} - 0.45  \quad  &13\lesssim \, \alpha \lesssim 30
\end{aligned}
\right.
 \eeq 
 where the first range $0.1 \lesssim \alpha \lesssim 7$ is corresponding with good approximation to all of the shapes of the 
 power spectrum analysed in Section~\ref{average}, suggesting that the other two ranges are suppressed by the smoothing.
 They describes energy density profiles which are very sharp around $\hr_m$ where the threshold is computed, and therefore 
 such profiles are smoothed by the pressure gradients, as we have seen in the right panel of Figure~\ref{rho_alpha}, suppressing 
 the values $\delta_c \gtrsim0.6$.

 This interpretation is enforced when the threshold is computed at the non linear horizon crossing, which is well described by 
 another fit, again divided into three different regimes.
 \beq \label{delta_cH_fit}
 \delta_c (t_H) \simeq
 \left\{
\begin{aligned}
 &\alpha^{0.125} - 0.05  \quad \quad &0.1\lesssim\alpha \lesssim 3 \\ 
&\alpha^{0.06} + 0.025  \quad\quad  & 3 \lesssim \,\alpha \lesssim 8 \\  
& \quad \quad 1.15  \quad  &\alpha \gtrsim 8
\end{aligned}
\right.
 \eeq 
 Here the first regime of~\eqref{delta_c_fit} is basically splitting into two different behaviors of the threshold computed at the 
 non linear horizon crossing time, while the second and the third regimes of~\eqref{delta_c_fit}, corresponding to $\delta_c \gtrsim0.6$
 computed at superhorizon scales, saturate to an almost constant value of the threshold when is computed at $t_H$.
  
 The right panel of Figure~\ref{fig:delta_c} shows that the ratio between $\delta_c(t_H)$ and $\delta_c(t_i)$, where one can 
 distinguish two different regimes: the first one, when  $\alpha \lesssim 3$, is corresponding to the increasing behavior 
 of this ratio, and explains the first regime of~\eqref{delta_cH_fit}. The second regime, when $\alpha \gtrsim 3$, has a 
 decreasing behavior of the ratio between the two thresholds, corresponding to the second and third regime
 of~\eqref{delta_cH_fit}, which can be distinguished in the right panel of Figure~\ref{peak_c}. 
 
 The lower  ($\alpha \gtrsim 0.1$) and the upper  ($\alpha\lesssim 30$) boundaries of validity of the fit are given by the numerical simulations that are not able to handle  
 very extreme shapes  beyond these values. We are however neglecting only a range of $\alpha$ which is not significant as we are already close 
 enough to the limits of $\delta_c$.
  
 Finally we can observe that the difference between the threshold computed at the non linear horizon crossing and the linearly 
 extrapolated one is an almost constant numerical coefficient, varying between $1.7$ and $2$. This underlines the fact that the 
 threshold $\delta_c$ is a much more stable quantity than the local critical amplitude of the peak, and has to be preferred for 
 distinguishing between cosmological perturbations forming PBHs and the ones that are bouncing back into the expanding medium.


\section{Conclusions}
\label{conclusions}
\begin{figure}[t!]
 \vspace{0.2cm}
 \centering
   \includegraphics[width=0.47\textwidth]{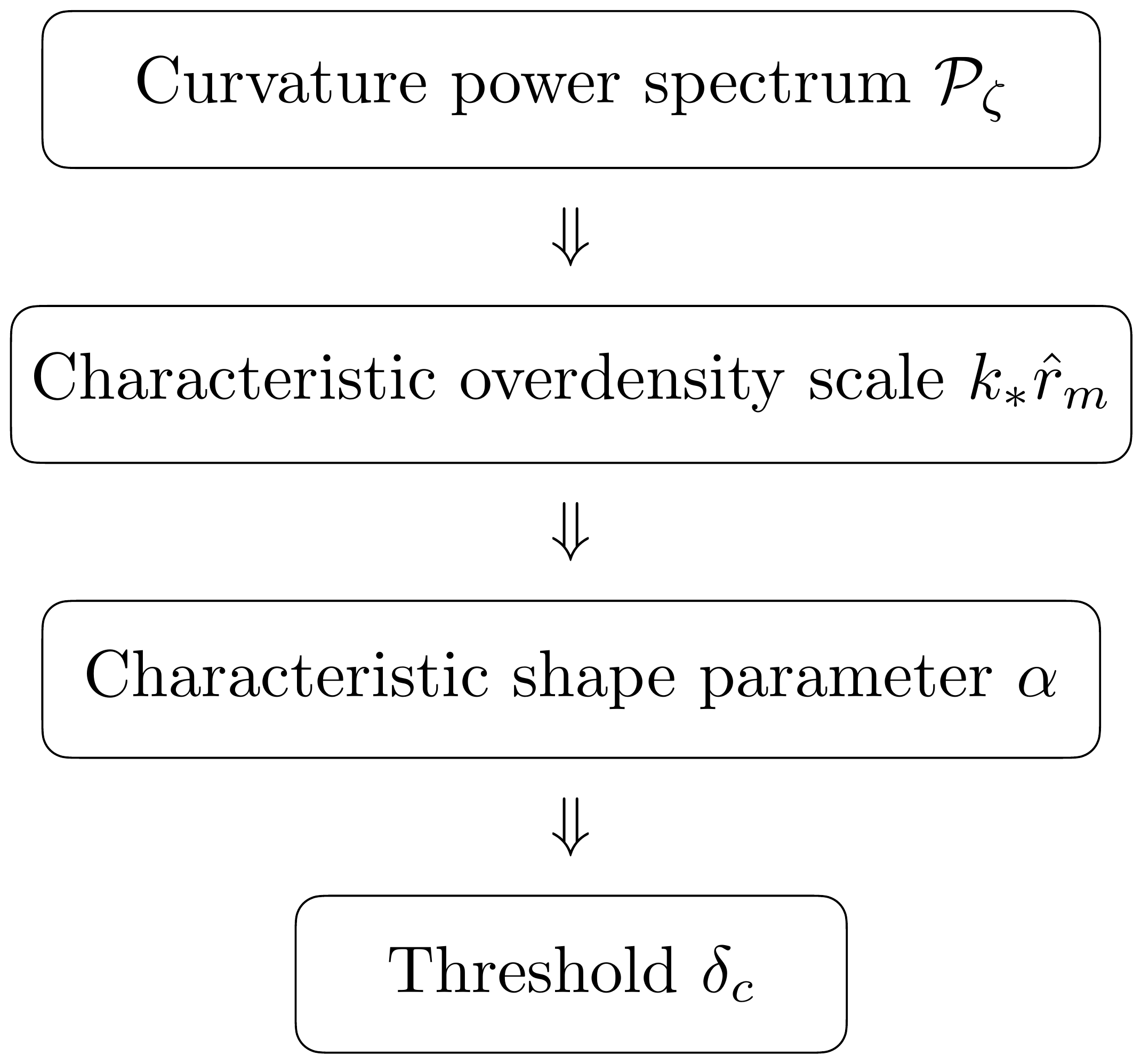} 
    \vspace{0.1cm}
  \caption{ This diagram summarises our prescription for computing the threshold $\delta_c$ starting from the power spectrum 
  of cosmological curvature perturbations ${\cal P}_\zeta$.  }
  \label{flow}
 \end{figure}

PBHs could have formed in the early universe from the collapse of cosmological perturbations at the horizon re-entry, 
provided that their amplitude is larger than a certain critical threshold. In this paper we have provided a simple analytical 
prescription, summarised in Fig. \ref{flow}, to compute the threshold of collapse for PBHs, embedding results coming 
from numerical simulations. 

From Gaussian curvature perturbations, one can compute the mean profile on superhorizon scales using peak theory and 
find the characteristic comoving scale of the perturbations from the given shape of the curvature power spectrum. From the 
computation of the profile shape parameter on superhorizon scales, one can determine the value of the threshold, also taking 
into account the effects of non-linearities arising at the cosmological horizon crossing fitted from  numerical simulations. In 
particular we stress that the thresholds calculated at horizon crossing differs by a factor of order two from the values traditionally 
adopted in the literature. 

By analysing different explicit examples of the curvature power spectrum, we have seen that in general the value of the threshold 
$\delta_c$ is larger for a monochromatic power spectrum, modelled by a Dirac delta, than for a broader shape which allows 
more modes to contribute to the collapse. The latter gives a broader and flatter profile of the compaction function describing a 
cosmological perturbation collapsing to form a PBH, corresponding to a lower value of $\delta_c$.  This allows, using 
\eqref{delta_c_fit}, to compute the threshold of PBHs measured on super-horizon scales by correctly identifying the shape parameter 
for a given curvature power spectrum, obtaining $0.4 \lesssim \delta_c \lesssim 0.6$. 

However, if the threshold is computed at the non linear horizon crossing time (i.e. around the time when they are really formed), 
the physical range of the threshold $\delta_c$ obtained from \eqref{delta_cH_fit} for all of the possible shapes of the power spectrum 
is $ 0.7 \lesssim \delta_c(t_H) \lesssim 1.15\,$. This might introduce a sizeable contribution in the calculation of the corresponding 
abundance of PBHs which is exponentially sensitive to the squared value of the threshold $\nu_c \equiv \delta_c/\sigma$, where $\sigma$ 
is the variance of the density field of cosmological perturbations. So far in the literature those have been computed on super horizon scales, 
which gives only the leading order computation of the abundance. Non linear effects, becoming important close to the horizon crossing, 
can give rise to corrections to the probability of collapse estimated on super horizon scale. The full computation of the abundance, however, 
would require knowing both $\delta_c$ and $\sigma$  at the non linear horizon crossing. In this work we have provided the first step in this 
direction by computing the threshold also at the exact horizon crossing time. The corresponding computation of the variance would however 
require a dedicated analysis of the non linear transfer function, which is beyond the aim of this work.

\vspace{-1.0cm}
\section*{Acknowledgments}
We thank Silvio Bonometto, Cristiano Germani, John Miller and Sam Young for useful comments.
I.M. is supported by the ``Mar\'ia de Maeztu'' Units of Excellence program MDM-2016-0692 and the Spanish Research 
State Agency. I.M. thanks the Department of Theoretical Physics of the University of Geneva for financial support and 
hospitality, and CERN for financial support and hospitality during the final completion of this paper.  V.DL., G.F. and A.R. 
are supported by the Swiss National Science Foundation (SNSF), project {\sl The Non-Gaussian Universe and Cosmological 
Symmetries}, project number: 200020-178787. 



\end{document}